\documentclass[aps,twocolumn,amsmath,amssymb,superscriptaddress,prb,longbibliography]{revtex4-2}

\usepackage{graphicx}
\usepackage{dcolumn}
\usepackage{float}
\usepackage{amsmath}
\usepackage{lipsum} 
\usepackage{bm}
\usepackage{caption}
\usepackage{subcaption}
\usepackage{ragged2e}
\usepackage[normalem]{ulem}
\usepackage{xcolor}
\usepackage{ulem}

\usepackage[linktocpage=true,colorlinks=true,pdfborder={0 0 0},linkcolor=blue,citecolor=blue,filecolor=yellow,urlcolor=blue,bookmarks,pdfauthor={},]{hyperref}

\begin{document}

\preprint{APS/123-QED}

\title{Enhancing Coherence of Spin Centers in $p$-$n$ Diodes via Optimization Algorithms}

\author{Jonatan A. Posligua}
 \affiliation{Department of Physics and Astronomy, University of Iowa. Iowa City, IA, USA.} 
 \author{David E. Stewart}%
\affiliation{%
 Department of Mathematics, University of Iowa. Iowa City, IA, USA.
}
 \author{Denis R. Candido}%
 \email{denis-candido@uiowa.edu}
\affiliation{Department of Physics and Astronomy, University of Iowa. Iowa City, IA, USA.}
\affiliation{Applied Mathematical \& Computational Sciences, The University of Iowa, Iowa City, Iowa, USA}




\date{\today}

\begin{abstract}
Solid-state spin defects hold great promise as building blocks for various quantum technologies. Embedding spin centers in $p$-$n$ diodes under reverse bias has proved to be a powerful strategy to narrow the optical linewidth and increase spin coherence, while also enabling control of the photoluminescence wavelength via Stark shift. Given the multitude of parameters influencing spin centers in diodes (e.g., doping densities and profiles, temperature, bias voltage, spin center position), a question that has not yet been answered is: which set of these design parameters maximizes spin center coherence? In this work, we address this question by developing a scaled gradient descent optimization algorithm that minimizes the optical linewidth of spin centers by combining the numerical solution of a diode's Poisson equation with calculated charge noise from the non-depleted regions.
Our optimization is performed for both single- and multiple-parameter cases for divacancies in SiC $p$-$i$-$n$ diodes, including reverse-bias voltage, doping density and profile, and diode total length.  Importantly, the optimization is subject to realistic physical constraints, such as small operating bias voltages, avoidance of the dielectric breakdown regime and physical thresholds for doping density.
Additionally, due to the leakage current at reverse bias voltages, we develop a new formalism to investigate its influence on coherence. We show that the corresponding noise can be mitigated by implanting spin defects away from the diode's surfaces. Our work provides guidance on experimentally relevant diodes for hosting spin centers with the narrowest optical linewidths and longest coherence times.

\end{abstract}

\maketitle


\section{Introduction}

Spin center defects in solid-state systems have recently attracted interest for their potential to serve as different building blocks for quantum technologies~\cite{spinqubit1,spinqubit2,spinqubit3,spinqubit4}.
Their coherence time \cite{ScenterApp5,ScenterApp6} and operation at room temperature \cite{ScenterApp2,ScenterApp2a,ScenterApp2b}, together with their optical spin-state initialization and readouts \cite{ScenterApp1,ScenterApp4} promoted them as an alternative platform for quantum computing \cite{ScenterApp7,ScenterApp8}, quantum networking \cite{networking1,networking2,networking3,networking4}, single-photon quantum emitters  \cite{spinEM1,spinEM2,spinEM3}, and non-evasive sensing of electromagnetic fields \cite{sensing1,sensing2,sensing2.1}, temperature \cite{sensing3,sensing4}, 
and biological systems \cite{sensingBIO1,sensingBIO2,sensingBIO3,sensingBIO4}.
Despite their great potential, spin centers are susceptible to various noise sources that limit their performance. Environmental perturbations such as interaction with nearby nuclear spins~\cite{spin1,spin2}, charge and spin fluctuations~\cite{noiseMODE3,Candido1,rates,SpinCenter1,5rjw-ygrn,RevModPhys.97.021001}, and phononic disturbances~\cite{phonons1,phonons2} introduce undesired effects such as spectral diffusion \cite{noiseMODE1,phonons1,charge2}, which degrades coherence and photon indistinguishability \cite{noiseMODE2}. The shortening of the coherence times induced by noise~\cite{noiseMODE3} not only reduces the fidelity of quantum operations \cite{fidelity1,fidelity2}, but also constrains the sensitivity of spin-based sensing protocols via the decrease of both optical contrast and signal-to-noise ratio \cite{noiseMODE4,noiseMODE5}. As a result, suppressing and mitigating noise mechanisms are crucial for unlocking the full capabilities of spin centers in quantum technologies.

\begin{figure*}[t!]
    \centering
    \includegraphics[width=\textwidth]{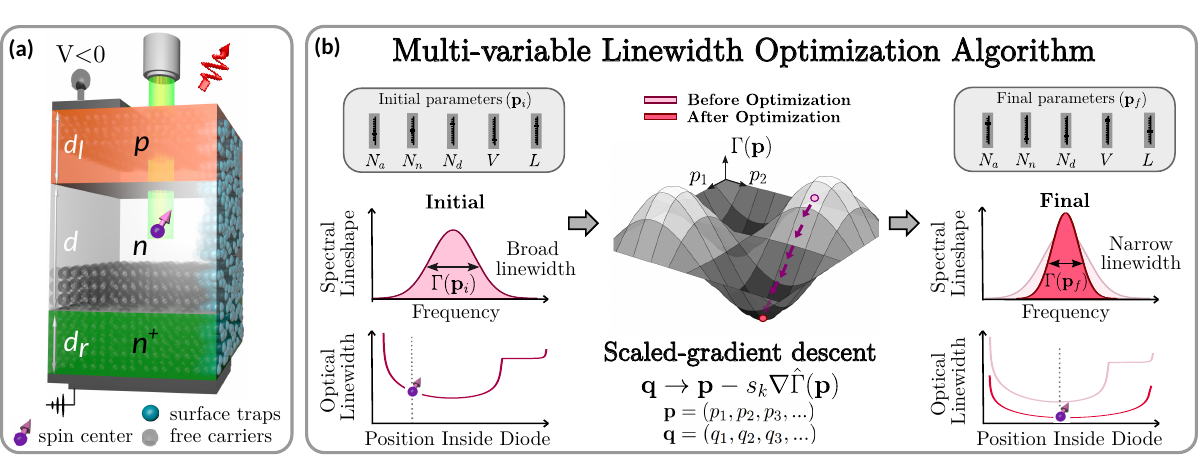}
    \caption{(a) Schematic of a 4H-SiC $pnn^+$ diode operated in the reverse bias regime with embedded spin centers, non-depleted charges, and surface traps responsible for the leakage current. (b) Schematic of the multi-variable linewidth optimization algorithm via the scale-gradient descent method. The linewidth is first calculated for an initial set of diode design parameters $\mathbf{p}$, and then autonomously minimized using a scaled gradient descent method. Iterations are performed until convergence of the linewidth is achieved, resulting in a final set of design parameter.}
    \label{FIG1}
\end{figure*}

Strategies for noise reduction include isotopically purified hosts~\cite{doi:10.1073/pnas.2121808119,Wolfowicz2021}, surface treatment~\cite{treatment1,treatment2,treatment3}, coating~\cite{coating1,coating2,passv4,coating3,sensingBIO4}, passivation~\cite{passv1,passv2,passv3}, as well as sample cooling \cite{cooling1,cooling2}; whereas noise mitigation can be achieved through schemes such as Dynamical Decoupling (DD) sequences \cite{DD1,DD2,DD3} and optimal control theory \cite{control3,control4}.
Recent and past proposals highlight $p$-$n$ diodes as good hosts for quantum information units such as spin centers ~\cite{Candido1,nvdiode1,nvdiode2,nvdiode3,nvdiode4,nvdiode5,nvdiode6}
and quantum dots \cite{qdots1,qdots3,qdots4} [See Fig.~\ref{FIG1}(a)]. For instance, Anderson et al. \cite{nvdiode1} observed a 50-fold narrowing of the optical linewidth of divacancy spin centers embedded in 4H-SiC $p$-$n$ junctions under reverse bias, with a Stark-induced shift exceeding 850 GHz. This was quantitatively explained by Candido et. al., \cite{Candido1} as a direct consequence of the increase of the charge depletion region around the spin centers' locations. Diodes have also demonstrated the electrical control of the emitted wavelength and charge states of quantum dots~\cite{qdots1,qdots4}.

Despite significant demonstrations of diodes as a key host for quantum units, it remains unknown how to engineer them to maximize coherence of embedded quantum units 
under experimentally relevant constraints, such as operating voltages, doping densities, and device dimensions.
Large reverse voltages reduce charge noise due to a large depletion layer, but it requires a voltage amplifier that introduces additional electric noise and degrades stability. High voltages also push diodes closer to the dielectric breakdown and increase leakage current and associated noise, causing higher power dissipation and thermal effects that can otherwise degrade spin center's performance and stability. Small voltages ($\lesssim 10$~V) are therefore technically and experimentally advantageous as they are generated with high precision, high quality, and with more stability.

To advance the use of diodes in quantum technologies, this work intends to fundamentally understand what are the sets of the diode's ``design parameters'' (i.e., bias voltage, doping densities and profile, diode total length) that maximize coherence of embedded spin centers while respecting relevant constraints.
Addressing these types of questions is crucial to achieving optimal performance in related quantum technologies. 
Most importantly, this requires a multidisciplinary and complementary approach involving optimization~\cite{OPT1} and machine learning tools~\cite{ML1} combined with expertise in solid-state physics and quantum information science. Recently, various works have tackled similar questions, e.g., physics-aware machine learning algorithms determining favorable gate voltage configurations in quantum dot hetero-structures \cite{decoherence1} and deep learning algorithms mitigating environmental noise experienced by qubits~\cite{decoherence5}. Optimization schemes have also been developed to determine diode parameters resulting in diodes with high power and low noise at THz frequencies~\cite{PrevOpt}.

In this work, we provide guidelines for diode characteristics and configuration that yield maximally coherent embedded quantum units. We find optimal diodes by developing a multi-variable constrained optimization algorithm~\cite{PrevOpt2,PrevOpt3,PrevOpt4} that minimizes optical linewidth of a spin center embedded in a $p$-$i$-$n$ with respect to the diode's design parameters, including bias voltage, doping densities and profile, and the diode's total length.
We build a general formalism that can be applied to diodes made of any material, whereas we focus our analysis on 4H-SiC diodes with embedded (hh) and (kk) di-vacancies [Fig.~\ref{FIG1}(a)] due to their exceptional spin center coherence~\cite{nvdiode1}, well-established material fabrication~\cite{4hsicFAB1,4hsicFAB2}, high breakdown voltages~\cite{SiC3,breakd2,breakINTRO3} and room temperature operation~\cite{4hsicTEMP1,4hsicTEMP2}
. Our formalism is schematically shown in Fig.~\ref{FIG1}(b). It consists of first solving the Poisson's equation for an initial set of diode parameters under reverse bias voltage. Charge noise originating from the non-depleted regions affecting the optical linewidth is then calculated using the free-carrier density profile ~\cite{Candido1}. Next, we apply our algorithm to optimize the optical linewidth with respect to the diode's design parameters via a scaled-gradient descent method~\cite{SGDO1,SGDO2} that takes our parameters toward minima regions in the parameter space. Most importantly, our algorithm incorporates different physical constraints such as both upper and lower thresholds on impurity densities, bounds on doped layer sizes, and avoidance of the dielectric breakdown~\cite{breakINTRO1,breakINTRO2,breakINTRO3}.

We first investigate linewidth optimization with respect to a single type of parameter, including voltage and all diode doping densities, separately. 
Optimal linewidth narrowing is achieved by increasing the reverse bias voltage and reducing the densities, corresponding, respectively, to a larger charge-depletion region around the spin center and fewer free carriers generating charge noise. Smaller intrinsic doping densities relative to extrinsic densities are shown to be favorable for noise reduction, in addition to enabling a large depletion region at low voltages and fine-tuning control over the depletion length and Stark shift. Diodes 
Secondly, multi-parameter optimization with respect to all the lengths of the $p$, $i$, and $n$ regions is achieved by having larger intrinsic regions than the extrinsic $p$ and $n$ regions. 
Furthermore, as it is often required to operate diodes across different lengths and voltage regimes, we perform multi-parameter optimization with respect to voltage and densities for three fixed diode lengths of $0.1~\mu$m, $1~\mu$m, and $10~\mu$m. Similarly, optimized linewidth is achieved by increasing the voltage and reducing the densities. As the diode's overall length decreases, we observe that physical constraints of our systems limit the optimal parameters, such as lower thresholds on doping densities and maximum applied voltages. 

Furthermore, we develop a formalism to calculate the impact of the electromagnetic noise due to the leakage current on the spin center coherence \cite{LEAKS7,LEAKS10,LEAKS13,IVcurve}.
This is important since increasing the voltage — and consequently the leakage current — effectively suppresses the linewidth, but could introduce further decoherence processes to our spin centers. We show that leakage current noise strongly depends on the local density of surface impurities, which decreases as a function of the spin center's depth from the diode's surface. While large values for the corresponding linewidth are obtained for shallow spin centers, this can be mitigated by placing the defects far from the surface. 

Our paper is organized as follows. In Sec. \ref{Diode}, we introduce the diode theory for our problem. Sec. \ref{Noise} introduces the linewidth formalism as well as its relation to the noise arising from carrier fluctuations in diodes. Sec. \ref{Opt} outlines our scaled-gradient descent optimization formalism, and the mathematical models used to describe the physical constraints of the diode system. In Sec. \ref{Results}, we present and discuss our optimization results with respect to reverse bias voltage, doping densities, lengths of the diode's layers, and for different experimentally-relevant length regimes. Finally, also as part of Sec. \ref{Results}, we assess the behavior of the diode's design parameters when subject to the physical constraint associated with the avoidance of dielectric breakdown.

\section{Diode Theory}
\label{Diode}

\vspace{-1em}
In our work, spin centers in 4H-SiC are assumed to be implanted within a diode configuration under reverse bias voltage $V<0$. Our $p$-$n$-$n^+$ diode is grown along the $z$-direction, and consists of a $p$-doped ($p$), intrinsic or lightly $n$-doped ($n$), and $n$-doped regions ($n^+$), as illustrated in Fig.~\ref{FIG1}(a). The $p$-region ($-d_l<z<0$) has acceptor density $N_{a}$, the $n$ region ($0<z<d$) has light donor density $N_n$ and the $n^+$ region ($d<z<d+d_r$) has donor density $N_d$. For simplicity, we assume that the change in potential along the diode's growth direction is much larger than the variations along the remaining dimensions. Therefore, the electrostatic properties of the diode are determined from Poisson's equation,

\begin{equation}
\frac{\partial^{2}\phi\left(z\right)}{\partial z^{2}} = -\frac{\rho\left(z\right)}{\epsilon},
\label{EQ8}
\end{equation}

\noindent where $\phi(z)$ is the electrostatic potential, $\rho(z)$ the charge density as a function
of position $z$, $\epsilon$ the material's permittivity
constant defined as $\epsilon=\kappa\epsilon_{0}$ with $\kappa\approx9.66$ being
the 4H-SiC dielectric constant, and $\epsilon_{0}=8.99\times10^{-12}$ F/m the permittivity of free space.
The total charge density $\rho(z)$ has a contribution from the spatial background charge arising from both ionized impurities [$N_A(z)$ for acceptors, $N_N(z)$ and $N_D(z)$ for donors] and electron and hole free carriers [$\rho_c(z)$], yielding
\begin{widetext}
\begin{equation}
\begin{aligned}
\rho(z)=q \Bigg[
& \overset{\equiv N_{N}(z)}{\overbrace{\left[\frac{N_{n}}{2e^{\frac{\mu_{r}+q\phi(z)-\epsilon_{d}}{k_{B}T}}+1}\right] \Theta(d-z)\,\Theta(z)}}- \overset{\equiv N_{A}(z)}{\overbrace{\left[\frac{N_{a}}{2e^{\frac{-\mu_{l}-q\phi(z)+\epsilon_{a}}{k_{B}T}}+1}\right] \Theta(z+d_{l})\,\Theta(-z)}} \\
&+ 
\overset{\equiv N_{D}(z)}{\overbrace{\left[\frac{N_{d}}{2e^{\frac{\mu_{r}+q\phi(z)-\epsilon_{d}}{k_{B}T}}+1}\right] \Theta(z-d)\,\Theta(d+d_{r}-z)}}
+ 
\overset{\equiv\rho_{c}(z)}{\overbrace{n_{i}(T,E_{g})  \left( e^{\frac{-\mu_{l} - q\phi(z) + \epsilon_{i}}{k_{B}T}} - e^{\frac{\mu_{r} + q\phi(z) - \epsilon_{i}}{k_{B}T}} \right)}}
\Bigg],
\label{EQ8_modified}
\end{aligned}
\end{equation}
\end{widetext}
where $q=e=1.6\times10^{-19}$ C is the fundamental electric charge, $k_B$ the Boltzmann's constant, $T$ the system's temperature. $\epsilon_a$ ($\epsilon_d$) is the acceptor (donor) energy level, $\Theta$ the Heaviside step-function, $\mu_{l}$ ($\mu_{r}$) the quasi-Fermi level evaluated at $z=-d_l$ ($z=d_r$). \( \epsilon_i \equiv \frac{\epsilon_c + \epsilon_v}{2} + \frac{k_B T}{2} \ln\left(\frac{P_v(T)}{N_c(T)}\right) \) is the intrinsic Fermi energy with $N_{c}\left(T\right)\equiv\frac{1}{4}\left(\frac{2m_{c}k_{B}T}{\pi\hbar^{2}}\right)^{3/2}$, $P_{v}\left(T\right)\equiv\frac{1}{4}\left(\frac{2m_{v}k_{B}T}{\pi\hbar^{2}}\right)^{3/2}$, and $n_{i}(T, E_{g}) \equiv \sqrt{N_{c}(T)P_{v}(T)}e^{-\frac{E_{g}}{2k_{B}T}}$
, where  $m_{c}$ ($m_{v}$) is the electron (hole) mass, $\hbar$ is Planck's constant, and $E_g\equiv\epsilon_c - \epsilon_v$ the band-gap energy with $\epsilon_c$  ($\epsilon_v$) as the conduction (valence) band energy.

\begin{figure*}[t!]
    \centering
    \includegraphics[width=1.0\textwidth, clip]{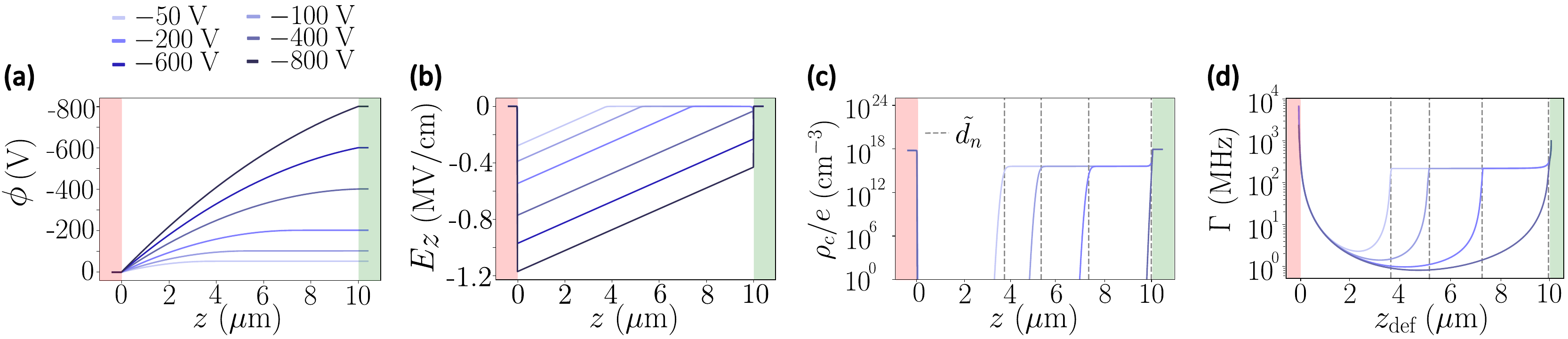}
    \caption{Electrostatic properties of a 4H-SiC $pnn^+$ diode: (a) Electric potential along the diode's $z$-direction calculated for reverse bias voltages ranging from $-50$ to $-800$ $\mathrm{V}$. (b) Same for the electric field. (c) Total charge carrier density as a function of position within the diode's $z$-direction calculated using the solution to Poisson's equation to evaluate $\rho_c$ in Eq. (\ref{EQ8_modified}) for reverse voltages ranging from $-50$ to $-400$ $\mathrm{V}$. The depletion boundaries $\tilde{d_n}$ are shown for each reverse bias value. (d) Same for the optical linewidth as a function of the spin center's position within the diode's $n$-region (i.e., $z_{\text{def}}$). The doping densities used to generate all the sub-figures are $N_a=$ $7\times10^{18}\mathrm{cm^{-3}}$, $N_n=$ $4\times10^{15}\mathrm{cm^{-3}}$, $N_d=$ $1.01\times10^{19}\mathrm{cm^{-3}}$; whereas the lengths of the diode's layers are $d_l=d_r=$ $0.4$ $\mathrm{\mu m}$ and $d=$ $10$ $\mathrm{\mu m}$. The temperature is $T=300$ K.}
    \label{FIG2}
\end{figure*}


When a reverse voltage is applied across the diode, the Fermi level is no longer constant. The difference between the Fermi levels evaluated at the diode's ends is $eV$ \cite{Poisson4,Poisson5}, thus $\mu_r=\mu_l+eV$. The boundary conditions used to solve Eq. (\ref{EQ8_modified}) are $\phi(z=-d_l)=0$ and $\phi(z=d_r)=\phi_\infty$, where $\phi_\infty$ is the built-in potential across the diode for any bias voltage \cite{Candido1}. In our calculations, the values of $\mu_l$ and $\phi_\infty$ are obtained by ensuring charge neutrality at $z=-d_l$ and $z=d_r$, respectively (See Appendix \ref{AppA}). Eq. (\ref{EQ8_modified}) is a non-linear second-order differential equation for $\phi(z)$, which we solve numerically via the Newton-Raphson method~\cite{Poisson1}. Convergence issues are present in this differential equation due to large or small values of the arguments of the exponential functions. To properly address this situation, we enclose all the arguments of the exponential functions within a regulating Sigmoid function \cite{solver}, such that, at any iteration of the Newton-Raphson method, the value of the exponential function never exceeds the upper and lower bounds of the numerical tolerance. (See Appendix \ref{AppB}).

Fig. \ref{FIG2}(a) shows the solution to Eq. (\ref{EQ8_modified}) for different values of $V$, while Fig. \ref{FIG2}(b) shows the electric field for the same bias voltages of Fig. \ref{FIG2}(a). Moreover, Fig. \ref{FIG2}(c) presents the carrier density obtained from Eq. (\ref{EQ8_modified}) for bias voltages up to $-400$ V. For voltages smaller than $ V_c \approx -\frac{e d^2}{2\epsilon} \left( \frac{N_n (N_a + N_n)}{N_a} \right) \approx -370 $ V \cite{Candido1}, the lightly-doped $n$-region is fully depleted and the charge noise coming nearly insulating carrier vanishes.




\section{Modeling the Optical linewidth}
\label{Noise}

In this section, we outline how to quantitatively obtain the optical linewidth from the emission spectra of spin centers embedded in diodes. In diodes, charged localized states and both majority and minority carriers display fluctuations in their occupation and positions caused by thermal excitation and laser illumination in photoluminescence-based protocols~\cite{fluctuations1,fluctuations2,fluctuations3}. Fluctuations in the charge and current densities give rise to fluctuating electric and magnetic fields \cite{fluctuations4,fluctuations4b,fluctuations5}, leading to electromagnetic noise \cite{FDT1,FDT2,FDT3}
and decoherence mechanisms manifested as optical linewidth \cite{linewidth6,lineshape1}. 
For quantum emitters, narrowest linewidth is crucial to ensure the proper operation of emitters with well-defined transition frequencies 
\cite{linewidth3,linewidth4,linewidth5} and to enable photon indistinguishability studies crucial for quantum computing and networking.
We first introduce the general formalism for calculating the optical linewidth due to different types of noise in our diodes (Sec. \ref{NoiseA}). We then calculate the linewidth arising from the charge noise due to the fluctuation of majority charge carriers within the non-depleted diode's regions, following a similar approach to Ref. \cite{Candido1} (Sec. \ref{NoiseB}). Finally, Sec.~\ref{NoiseC} presents a new model for the linewidth broadening associated with leakage current, which we attribute to the generation of electric current \cite{LEAKS8,LEAKS5} within the diode's depletion region near the surface \cite{SURFACE_LEAK,SURFACE_LEAK2,rates}.

\subsection{Optical Linewidth}
\label{NoiseA}

To properly characterize the optical linewidth, we first obtain the line-shape describing the emission spectra of spin centers when subject to noise producing fluctuations in the spin frequencies, $\delta\,\omega(t)$. For this end, we apply Kubo's line-shape theory \cite{lineshape1} for a random Gaussian process, which gives the line-shape as a function of frequency as 
\begin{equation}
I(\omega) = \frac{1}{2\pi} \int_{-\infty}^{\infty} dt\, e^{-i\omega t} \mathcal{C}(t),
\label{optlnwth1}
\end{equation}
\noindent where $\mathcal{C}(t) = \exp\left[ -\frac{1}{2} \int_0^t dt' \int_0^t dt''\, \Pi(t', t'') \right]$ is the coherence function, with $\Pi(t', t'') = \left\langle \delta\omega(t') \delta\omega(t'') \right\rangle$ as the two-point time correlation function of the frequency fluctuations. Here, these fluctuations depend on the microscopic origin of the noise and on the spin Hamiltonian. For noise invariant under time translation $\Pi(t'',t')=\Pi(\tau=t''-t')$, the correlation can be characterized by the spectral noise density (SND) $S\,(\omega)$ via 
\begin{equation}
S\,(\omega)=\int_{-\infty}^{\infty}e^{-i\omega\tau}\,\Pi\,(\tau).
\label{SND_corr}
\end{equation}

\noindent Finally, the linewidth $\Gamma$ is given by the full width at half maximum (FWHM) of $I\,(\omega)$ in Eq. (\ref{optlnwth1}).

\subsection{Optical Linewidth due to non-depleted region (majority carriers)}
\label{NoiseB}

Non-depleted regions in the diode have majority carriers, whose fluctuations of their positions cause charge (electric) noise characterized by fluctuation of the electric field. These fluctuations are mainly caused by diffusion of charges across the different donors and by optical excitation during photoluminescence-based protocols. The associated correlated function $\Pi_E(\tau)$ decays exponentially in time ~\cite{noiseASUMP1,noiseASUMP2}, with the time scale of the fluctuations being much smaller than the system's coherence time (i.e., slow noise regime)~\cite{noiseASUMP3,nvdiode1}. This leads to a Gaussian $I(\omega)$ line-shape that is seen in corresponding experimental works~\cite{lineshape2}. Then, one can compute the line-shape using Eq. (\ref{optlnwth1}) and find its linewidth (See Appendix \ref{AppLNW}) as
\begin{equation}
\Gamma_E = \alpha\sqrt{\Pi_E\,(\tau=0)}\,.
\label{lnwdth14}
\end{equation}
Here, $\Pi_E\,(\tau=0)\equiv \frac{1}{2\pi}\int_{-\infty}^{\infty} S_{E_z}(\omega) \,d\omega\,$ with $S_{E_z}\,(\omega)$ denoting the spectral electric noise density for the $z$-component of the electric field, and $\alpha\equiv \sqrt{\frac{2 \ln 2}{\pi}}\,\frac{\mu_z}{ \hbar}\,$, with $\mu_z$ as the dipole moment along the diode's growth axis. Note that the square root of the correlation function evaluated at $\tau=0$ is nothing other than the standard deviation of the electric field, i.e., $|\delta E_z|\equiv\sqrt{\Pi_E\,(\tau=0)}$. We further assume isotropic noise, i.e., $|\delta E_z| = \frac{|\delta \mathbf{E}|}{\sqrt{3}}$, with $|\delta \mathbf{E}|$ obtained using the formalism in Ref. \cite{Candido1}, which considers an effective fluctuating dipole density produced by continuous trapping-de-trapping of free carriers in the vicinity of dopants $N_a$, $N_n$, and $N_d$. We present the relevant equations for the electric field fluctuations $|\delta E_z|$ in Appendix \ref{AppC1}. These expressions determine $\Gamma_e$ as a function of the diode's design parameters and the spin center's position, $z_{\text{def}}$. Fig.~\ref{FIG3}(d) uses the calculated depletion lengths from Fig.~\ref{FIG3}(c) to determine the optical linewidth from the equations presented in Appendix \ref{AppC1}. Since we place our spin centers in the lightly-doped $n$-region (i.e., the region with smaller concentration of fluctuators), we restrict the position dependence of $\Gamma$ only for $0<z_{\text{def}}<d$.

\subsection{Optical Linewidth due to leakage current}
\label{NoiseC}

Within the diode's depletion region, the near-absence of majority carriers is offset by a finite population of free carriers generated via stochastic generation-recombination (G-R) processes \cite{GTAT,LEAKS5,LEAKS8}. In wide-bandgap 4H-SiC, the contribution from minority carrier diffusion is negligible; instead, the reverse leakage current is dominated by field-enhanced generation \cite{thermal1,HURKX,field_enhancement} within the diode's depletion region—a process significantly amplified by the Poole-Frenkel (PF) effect and trap-assisted tunneling \cite{HURKX,tat1,tat2}. This microscopic dynamics is primarily mediated by localized surface impurities and trap states \cite{LEAKS5,LEAKS8,SURFACE_LEAK,SURFACE_LEAK2}, where high capture and emission rates, coupled with strong local electric fields, lead to efficient carrier collection and macroscopically measurable leakage currents.  Fig.~\ref{FIG1}(a) shows the presence of the trap states, which are localized near the diode's surface ($x=0$), and extend up to a depth $x=D<x_{\text{def}}$. The density and occupation of trap states depend on the material's termination and band bending~\cite{LEAKS8,sensingBIO4,passv4}.
The generated current and random trapping and de-trapping of these carriers generate electromagnetic noise~\cite{SpinCenter1,rates}, broadening the optical linewidth of nearby spin centers. It has been experimentally verified (as well as theoretically simulated) that the leakage current in diodes increases with the magnitude of the applied reverse bias \cite{leak1,leak2,leak3,leak4}. As increasing reverse voltages also suppress the noise via depletion of the majority carriers, it is crucial to understand and assess this trade off. To model the electric noise arising from the leakage current, we assume that the SND is produced by fluctuating dipole charges near the diode's surface \cite{rates,Candido1}, whereas for the magnetic noise, we describe it by Johnson-Nyquist magnetic noise \cite{JohnsonNyquist1,JohnsonNyquist2} (see Appendix \ref{AppLNW}).
In Appendix \ref{AppLNW}, we derive the optical linewidth for both the electric ($\Gamma'_E$) and magnetic ($\Gamma'_B$) noise due to the leakage current,
\begin{align}
  \Gamma_{E}'(x_{\text{def}}) &= \alpha\sqrt{\Pi'_{E}\,(\tau =0,x_{\text{def}})}, \label{eq:lwdth_elec} \\[6pt]
  \Gamma_{B}'(x_{\text{def}}) &= \eta\,S_{B_z(3d)}(\omega=0,x_{\text{def}}), \label{eq:lwdth_mag}
\end{align}
where $\Pi'_E\,(\tau=0,x_{\text{def}})$ is the depth-dependent $(x_{\textrm{def}})$ standard deviation of the near-surface fluctuating electric field, $S_{B_z(\text{3D})}(\omega=0,x_{\text{def}})$ the SND corresponding to the near-surface magnetic noise, and $\eta$ a proportionality constant, all formally defined in Appendix \ref{AppLNW}. In App.~\ref{AppLKG} we present how the leakage current obtained from I-V curves dictates the linewidth contribution Eqs.~\ref{eq:lwdth_elec} and \ref{eq:lwdth_mag}.
We conclude that the linewidth due to leakage current is significantly suppressed by placing the spin centers sufficiently far away from the diode's surface (e.g., $x_{\text{def}}>100$ nm). Thus, as long as spin defects are not grown in proximity to the diode's surface, no further constraints are needed in our diode's design parameters, and only the noise due to majority carriers significantly contributes to the spectral broadening of our emitters. Therefore, in the upcoming sections, we only minimize the linewidth associated with the majority carriers present in the non-depleted diode regions.

\section{Linewidth Optimization}
\label{Opt}

In this section, we outline our optimization algorithm, which determines the optimal values of the optical linewidth  for spin centers' emission spectrum, while accounting for the relevant physical constraints of our system. The charge noise arising from non-depleted regions depends on the diode's design parameters $\mathbf{p}=(p_1,p_2,...)$ and position $z_{\text{def}}$ within the diode. The design parameters include doping densities, reverse bias voltage, and lengths of the diode's doping regions. 
The analytical dependence of the linewidth on $\mathbf{p}$ and $z_{\text{def}}$ (via Eq.~\ref{lnwdth14}) allows us to optimize this problem using classical 
multi-variable constraint optimization approaches. We use the scaled-gradient descent method~\cite{SGDO1,SGDO2}, subject to physical constraints on the diode's design parameters $\mathbf{p}$. This method iteratively updates the initial set of design parameters, $\mathbf{p}_i$, by an amount proportional to the negative gradient of the corresponding linewidth with respect to $\mathbf{p}$, i.e.,~\cite{GDO1,GDO2},
\begin{equation}
    \mathbf{p}_{i+1}\rightarrow \mathbf{p}_i -s_k \nabla_{\mathbf{p}} \Gamma(\mathbf{p},z_{\text{def}})\left.\right|_{\mathbf{p}=\mathbf{p}_i},
    \label{gradient}
\end{equation}
where $s_k$ is the learning rate. These partial derivatives are calculated via finite differences, with truncation errors in this method often resulting in stagnation and oscillations in gradient descent protocols \cite{trunc_error}. Thus, whenever standard finite differences yield non-resolvable magnitudes of partial derivatives, we replace them by the average of the derivatives from the most recent iterations (see Appendix \ref{AppD}), so that we obtain a non-vanishing change in $\Gamma$, similar to what is done in pattern search techniques \cite{search_opt,search_opt2}. Additionally, if the partial derivatives calculated via finite differences lead to abrupt changes in the design parameters at a given iteration, the step size dictated by $s_k$ self-adjusts to prevent overstepping in parameter space (see Appendix \ref{AppD}). Such a strategy is associated with adaptive step sizes \cite{adaptiveSIZE1,adaptiveSIZE2}, a commonly used technique in optimization protocols.

These iterations take an initial point in parameter space towards a local minimum in the surrounding region, illustrated schematically in Fig. \ref{FIG1}(b). We perform these iterations until the parameters converge, with the final set of design parameters $\mathbf{p}_N$ corresponding to a local minimum of the linewidth function. The optimization with respect to the spin position is obtained by finding the minimum of  $\Gamma(\mathbf{p},z_{\text{def}})$ with respect to $z_{\text{def}}$ at every iteration of our protocol. The inclusion of constraints on our parameters is important for studying technically advantageous diodes (e.g., low-voltage operating diodes), and for avoiding regions of parameter space that are experimentally unrealizable (e.g., ultra-low or ultra-high doping densities) or that drastically degrade the properties of either the diode or the spin center (e.g., dielectric breakdown and decoherence). Formally, the constraints are functions of the design parameters $\mathbf{p}$ and bound the parameter space. In our work, these are enforced via a linear projection of $\mathbf{p}$  onto the boundaries defined by the constraint functions (See Appendix \ref{AppD}). 

Finally, due to the different orders of magnitude of the parameters $\mathbf{p}$ (e.g., densities $\sim 10^{19}$ $\mathrm{cm^{-3}}$, and voltages $\sim 100$ $\mathrm{V}$), drastic differences on the gradient step sizes along different directions in the parameter space may be produced, favoring the change of some parameters over the others \cite{GDO1,GDO2}. To circumvent this scaling issue, we use a \textit{scaled}-gradient descent protocol~\cite{SGDO1,SGDO2} that incorporates a scaling matrix normalizing the components of the gradient vector, so that all design parameters $\mathbf{p}$ contribute more equally to the gradient step size in all directions of the parameter space. Fig. \ref{FIG4} presents a simplified flow chart of the steps followed in our optimization algorithm. Therein, we first obtain the optical linewidth $\Gamma$ via the solution of the Poisson equation and noise formalism for the initial design parameters. Then, we compute the gradient of $\Gamma$ to update the design parameters using scaled gradient descent. At each step we check if physical constraints are satisfied, and if so, output updated parameters. Such a process is repeated until convergence to a local minimum is achieved. Appendix \ref{AppD} presents the formalism of constraint-scaled gradient descent, as well as the step-by-step implementation of our optimization method.

\subsection{Relevant Constraints for Diodes}
\label{OptA}

In this section, we outline the relevant constraints on our diode's design parameters $\mathbf{p}$ entering our optimization algorithm. Formally, for our target function $\Gamma(\mathbf{p})$, the $j^{\text{th}}$ constraint ($j=1,...,N$) is a function $h_j$ of the design parameters $\textbf{p}$, such that $h_j(\mathbf{p})>0$ (see Appendix \ref{AppD} for more details). The first set of constraints corresponds to the thresholds for all the doping densities. As non-intentional doping densities of the insulator region are around $10^{14}$~cm$^{-3}$~\cite{unintentionalDOPE}, this is set as the lowest density value of the $n$-side. To avoid depletion of carriers near metallic contacts attached to the diode's boundaries, we set lower thresholds for the $p$ and $n^+$ sides as $10^{17}$~cm$^{-3}$\cite{metalSEMI1,metalSEMI2}. Upper doping thresholds are also required to avoid the change of original chemical composition of the diode's material~\cite{BNG3}. Our second constraint is the lower threshold of $100$~nm for the lengths of the diode-doping layers, which prevents our algorithm from entering negative lengths that are not physical, and small lengths that are not achievable with standard growth/doping techniques.

\begin{figure*}[ht]
    \centering
    \includegraphics[width=\textwidth]{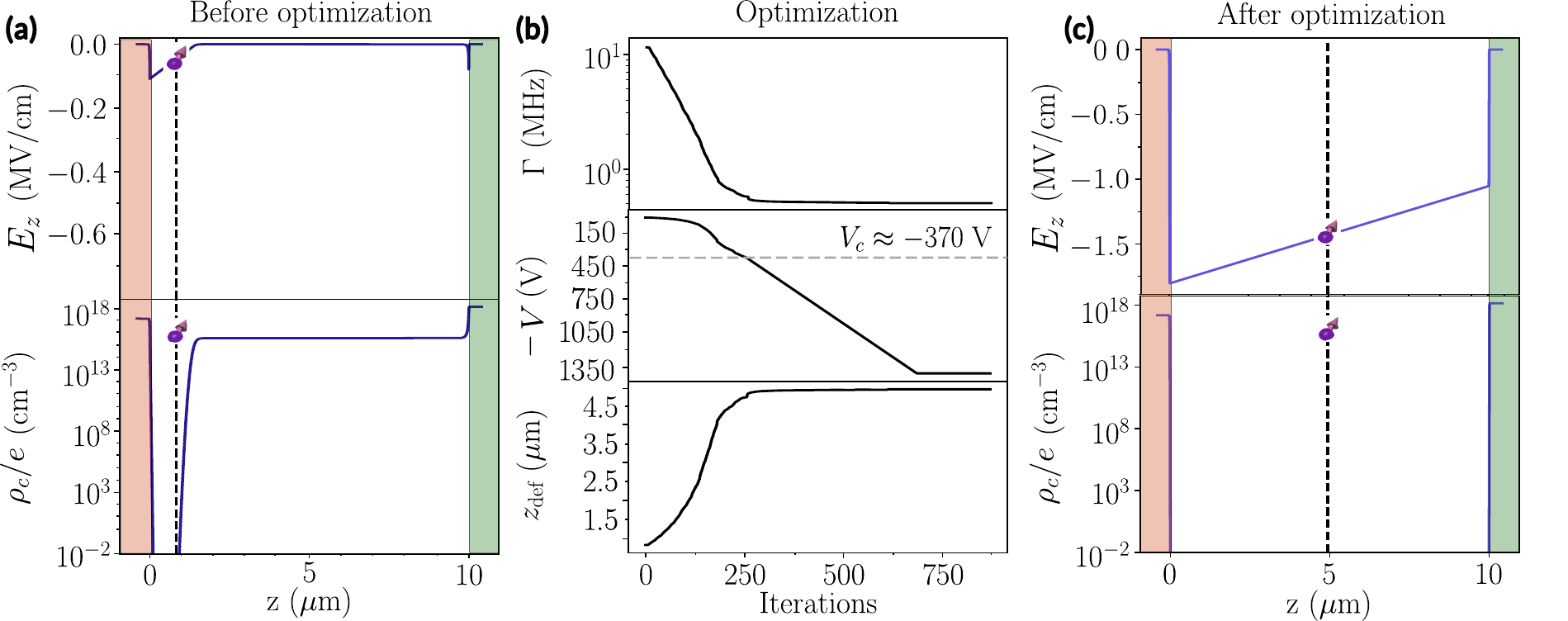}
    \caption{Single-parameter linewidth optimization with respect to the bias voltage. The initial design parameters are $N_a=7\times10^{18}$ $\mathrm{cm^{-3}}$, $N_n=4\times10^{15}$ $\mathrm{cm^{-3}}$, $N_d=1.01\times10^{19}$ $\mathrm{cm^{-3}}$, $T=300$ $\mathrm{K}$, $V=-5$ $\mathrm{V}$, $d_l=d_r=$ $0.4$ $\mathrm{\mu m}$ and $d=$ $10$ $\mathrm{\mu m}$. (a) Electric field and carrier density profiles for the initial parameters. (b) Optical linewidth, reverse bias voltage, and optimal spin center position as a function of iteration number. (c) Final electric field and carrier density profiles for the optimal design parameters. 
    }
    \label{FIG3}
\end{figure*}

Our third constraint is the upper bound on the electric field within the diode in order to avoid dielectric breakdown. Under large reverse bias voltages, the increased electric field magnitude enhances carrier generation from localized trap states \cite{LEAKS8} via the PF effect \cite{PF3,PF1,PF2} and trap-assisted tunneling \cite{tat1,tat2,GTAT}. These generated carriers are subsequently accelerated within the depletion region toward the diode's terminals. Dielectric breakdown happens when such carriers produce large currents that flow unimpeded in the diode \cite{LEAKS5}. This occurs when the electric field inside the diode is large enough to ionize the atoms of a given material \cite{breakINTRO0}, giving rise to secondary electrons and, subsequently, an electron avalanche \cite{breakINTRO1_a,breakINTRO1_b}. Such an avalanche increases electrical power, which leads to overheating and melting diodes. The magnitude of the electric field at which dielectric breakdown occurs can be calculated using the theory developed in Ref.~\cite{breakd1}, whose predictions are in agreement with the breakdown electric field reported in Silicon-based semiconductors \cite{breakd2}. The avoidance of the dielectric breakdown is enforced by requiring that for design parameters $\mathbf{p}$, the maximum value of the electric inside the diode is always smaller than $E_{BD}$, i.e., $\textrm{max}_z\{E_{z}(\mathbf{p},z)\}< E_{BD}$, with $E_{BD}\lesssim 2$~MV/cm~\cite{GODIGNON2024108347}.

\section{Results and discussion}
\label{Results}

In this section, we make use of our algorithm outlined in Sec.~\ref{Opt} for the optimization of the optical linewidth $\Gamma_e$ of di-vacancies in 4H-SiC due to solely charge noise from the non-depleted regions [Sec.~\ref{Noise}, Eq.~(\ref{lnwdth14})] while respecting the relevant physical constraints discussed in Sec.~\ref{OptA}. First, we perform single-type parameter optimizations with respect to voltage, diode densities, and the relative lengths of the $p$, $n$, and $n^+$ regions. Secondly, we perform multi-parameter optimization for all design parameters for different fixed diode total lengths ($0.1$~$\mu$m, $1$~$\mu$m, and $10$~$\mu$m). Finally, we demonstrate the robustness of our algorithm with respect to different initial conditions and the importance of respecting the physical constraints of our problem, such as avoiding dielectric breakdown, as discussed in Sec. \ref{OptA}. All of our optimizations enable us to understand the trends and regimes of parameters resulting in smaller linewidths of spin centers embedded in diodes.

\subsection{Optimization with respect to Bias Voltage}
\label{ResultsVOLT}

Here, we optimize $\Gamma$ with respect to the reverse bias voltage $V$, while keeping the other design parameters fixed. Our optimization is carried out for a spin center initially located at $z_{\text{def}}\approx0.85$~$\mu$m, and with $T=300$~K, $V=-5$~V, $N_a=7\times10^{18}$~cm$^{-3}$, $N_n=4\times10^{15}$~cm$^{-3}$, $N_d=1.01\times10^{19}$~cm$^{-3}$, $d_l=d_r=$ $0.4$ $\mathrm{\mu m}$, and $d=$ $10$ $\mathrm{\mu m}$. Fig.~\ref{FIG3}(a) illustrates both the electric field and the carrier density profiles prior to optimization. The corresponding linewidth is calculated via the formalism presented in Sec.~\ref{Noise}, yielding $\Gamma\approx11.1$~MHz at $z_{\text{def}}\approx0.85$~$\mu$m. We let the optimization algorithm run until convergence toward a local minimum is achieved. The upper panel in Fig.~\ref{FIG3}(b) shows the suppression of the linewidth as the iteration number increases, resulting in an overall twenty-fold suppression of the initial linewidth after convergence. The middle panel in Fig.~\ref{FIG3}(b) shows that such linewidth reduction is produced by increasing the magnitude of the reverse bias voltage $V$. Such an increase further depletes free carriers from the vicinity of the spin center, thus increasing the total depletion length of the lightly-doped $n$-region, ${d}_n (V)$, as \cite{Candido1}
\begin{figure*}[ht!]
\includegraphics[width=\textwidth]{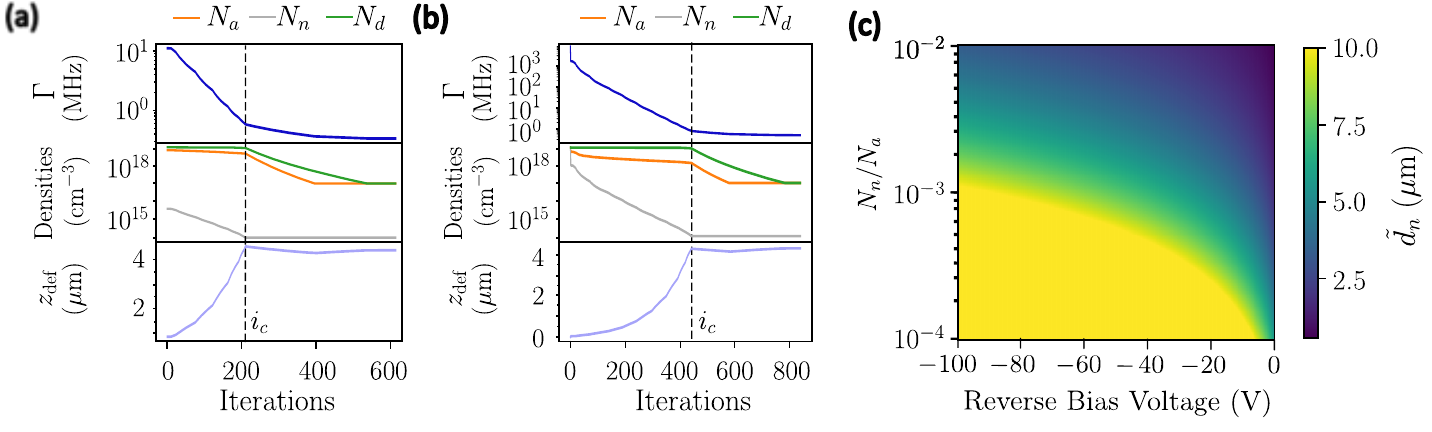}
\caption{(a) Single-type parameter linewidth optimization with respect to doping densities  for small bias voltage of $V=-5$~V, $N_n\ll N_a\approx N_d$, with $N_n=4\times10^{15}$ $\mathrm{cm^{-3}}$, $N_a=7\times10^{18}$ $\mathrm{cm^{-3}}$, and $N_d=1.01\times10^{19}$ $\mathrm{cm^{-3}}$, $T=300$~K, $d_l=d_r=$ $0.4$ $\mathrm{\mu m}$, $d=$ $10$ $\mathrm{\mu m}$. The upper panel shows the reduction in linewidth $\Gamma$ vs. iteration number, the lower panels show the corresponding doping densities and spin center position. (b) Same as (a), but for $N_a=N_n=N_d=10^{19}$~cm$^{-3}$ as initial conditions. (c) Density plot of depletion length within the intrinsic region ($d_n$) as a function of $N_n/N_a$ and $V$.}
\label{FIG4}
\end{figure*}
\begin{equation}
    {d}_n(V)=\sqrt{\frac{2\epsilon \phi_{\infty}(V)}{e}\frac{N_a /N_n}{N_a +N_n}},
    \label{EQ_dep_reg}
\end{equation} 
with $\phi_\infty \approx E_g /e - V$, which is plotted in Fig.~\ref{FIG2}(d). Charge noise arising from the free carriers is modeled as ${\delta\mathbf{E}\propto 1/(z-{d}_{n})^n}$, with $z-{d}_{n}$ being the distance of the spin center from the boundary of a non-depleted region, and $n$ a real number dependent on the character of the charge fluctuations~\cite{Candido1,SpinCenter1,rates}. This implies that the larger the voltage, the larger the depletion lengths, and hence the smaller the linewidth originating from electric noise. The upper panel of Fig.~\ref{FIG3}(b) shows that the reduction rate of the linewidth is greatest for the first $\sim250$ iterations. Around these iterations, $V \approx -370$~V, and Eq.~(\ref{EQ_dep_reg}) shows that the lightly-doped $n$-region is fully depleted, and so is the major contribution of the noise. Fig.~\ref{FIG3}(c) shows both the electric field and free-carrier charge density profiles across the diode after convergence, confirming the absence of free carriers in the $n$-region, and the larger electric fields compared to Fig.~\ref{FIG3}(a) due to the longer ionized doping region.

Further decreases in voltage deplete the $n$ and $p$ regions more, but because of the large distance to the carrier fluctuations and the small associated depletion length per voltage ($\sim 1$~nm/100~V~\cite {Candido1}), the corresponding linewidth decrease is tiny for iterations past $\sim 250$.
This is corroborated by experimental studies carried out by Zeledon et. al. \cite{nvdiode6}, which also demonstrated that the most prominent decrease in the optical linewidth occurs for voltages in which the diode's $n$-region becomes fully depleted. In principle, the optimization would continue until the voltage reaches the threshold $(V_{BD})$ at which electric breakdown is triggered. For the diode of Fig.~\ref {FIG3}, $E_{BD}\approx 1.9$ MV/cm giving $V_{BD}\approx E_{BD} d=1900$~V. Note that this is very far from the converged voltages of $\approx  -1350$~V, indicating that we are still far from the dielectric breakdown regime. This happens because once the $n$-region is fully depleted, the change in $\Gamma$ with respect to $V$ becomes very small in magnitude compared to $\Gamma$, making our code unable to resolve such tiny changes.
The optimization results near the dielectric breakdown regime are further explored in Sec. \ref{ResultsB}. Moreover, the lower panel of Fig.~\ref{FIG3}(b) shows that the optimal spin center position occurs near the middle of the diode (i.e., $z_{\text{def}}\approx d/2=5~\mu$m), which is equally far away from the charge fluctuators of both $p$ and $n^+$ regions. This is corroborated by Fig. \ref{FIG2}(d), where the spatial profile becomes symmetric as $V\rightarrow V_c$ for $N_a\approx N_d\gg N_n$.

\subsection{Optimization with respect to the diode doping densities for small voltages}
\label{ResultsDOPE}

Here, we investigate the optimization of the linewidth $\Gamma$ with respect to all the diode doping densities (i.e., $N_a$, $N_n$, and $N_d$) for small voltage $V=-5$~V, initial spin center position $z \approx 0.85$~$\mu$m, and $T=300$ $\mathrm{K}$, $d_l=d_r=$ $0.4$ $\mathrm{\mu m}$, $d=$ $10$ $\mathrm{\mu m}$. Fig.~\ref{FIG4}(a) shows the evolution of the linewidth (upper panel), diode densities (middle panel), and spin center position (lower panel) vs iteration number for initial parameters $N_n=4\times10^{15}$ $\mathrm{cm^{-3}}$, $N_a=7\times10^{18}$ $\mathrm{cm^{-3}}$, and $N_d=1.01\times10^{19}$ $\mathrm{cm^{-3}}$ (i.e., $N_n\ll N_a\approx  N_d$). A decrease in all doping densities reduced the initial linewidth by about a factor of thirty after the convergence of the algorithm, resulting in a sub-MHz final linewidth, similar to the ones obtained with applied high bias voltage of $\approx -1750$~V in Fig.~\ref{FIG3}. Interestingly, this demonstrates that significant charge noise mitigation in dioes can also be achieved without high-voltage operating setups.

To understand how this was achived, we separate the discussions before and after critical iteration $i_c\approx 200$. Before $i_c$, $\Gamma$ decreases mainly due to reducing both $N_n$ and $N_a$ more than $N_d$. For our initial condition, the $n$-region is not fully depleted, and the optimization code diminishes $N_n$ to increase the depletion length of the lightly-doped $n$-region, as $d_n\propto N_n^{-1/2}$ follows from Eq.~(\ref{EQC1_4}) for $N_a\gg N_n$. Additionally, because our spin center is initially closer to the $p$-region, decreasing the density of the lightly doped $n$ and $p$-regions reduces the density of the leading carriers that generate charge noise. $N_n$ stops decreasing at iteration $i_c$ as it has reached the lower threshold constraint for the densities of non-intentional doping $N_{crit}\approx10^{14}$~cm$^{-3}$ \cite{unintentionalDOPE} discussed in Sec.~\ref{OptA}. Finally, the spin center position is shifted towards the middle of the diode, further reducing noise from the non-depleted $p$ and $n$ regions.
For density values beyond iteration $i_c$, our diode presents a fully depleted intrinsic region, and thus charge noise comes from the heavily doped $p$ and $n$ regions. Therefore, optimization makes $N_d$ begin, and $N_a$ continues to decrease, further diminishing the density of charge fluctuators. This happens until we reach the constraint on the lower density threshold at $N_{a,crit}=N_{d,crit}=10^{17}$~cm$^{-3}$, which was set to avoid the formation of large depletion regions near the metallic contacts of the diode \cite{metalSEMI1,metalSEMI2} (See Sec.~\ref{OptA}). The lower panel in Fig. \ref{FIG4}(a) shows that the final optimal position occurs at $z_{\text{def}}\approx d/2$, which corresponds to an equally apart distance from fluctuators, similar to Sec. \ref{ResultsVOLT}.

We further test the robustness of our approach to different parameter initial values,
with corresponding results in Fig. \ref{FIG4}(b) for $N_a=N_n=N_d=1\times 10^{19}$ $\mathrm{cm^{-3}}$, and other parameters the same as the ones in Fig. \ref{FIG4}(a). The upper panel of Fig.~\ref{FIG4}(b) shows that, within the first $\approx20$ iterations, the linewidth is 2--3 orders of magnitude larger than that of Fig. \ref{FIG4}(a). This is due to the initial value of $N_n$ being significantly larger in this trial, leading to a higher density of fluctuators in the $n$-region, where the spin center is initially located. The behavior of the densities vs iteration number follows the same trend as in Fig. \ref{FIG4}(a), i.e., $N_a$ and $N_n$ decrease while $N_d$ remains unchanged up to iteration $i_c\sim 440$, after which $N_d$ also begins to decrease. Ultimately, all densities reach their respective lower thresholds at the final iterations, and optimal position converge towards the center of the diode. The analysis of the final linewidth confirms that $\Gamma$ has converged to the same local minimum as that from Fig. \ref{FIG4}(a).

In Fig. \ref{FIG4}(c) we plot the depletion length of the $n$-region, $d_n(V)$ [Eq. (\ref{EQ_dep_reg})], for different values of $V$ and $N_n/N_a$, with $N_{a}=1\times 10^{18}$ $\mathrm{cm^{-3}}$ and $1\times 10^{15}\,\mathrm{cm^{-3}}\le N_n \le 1\times 10^{18}\,\mathrm{cm^{-3}}$. Interestingly, we observe that the regime obtained by the optimization results, $N_n\ll N_a$, is also relevant for controlling the depletion length with slight variations of small bias voltage. In summary, our results indicate that when operating with small bias voltages, control over impurity growth can reduce the linewidth by more than one order of magnitude by decreasing the densities of $p$, $n$, and $n^+$ to their respective lower-density thresholds. Nevertheless, this also leads to smaller electric fields experienced by the spin centers due the smaller background charge from ionized doping regions.


\begin{figure}
\includegraphics[width=0.48\textwidth]{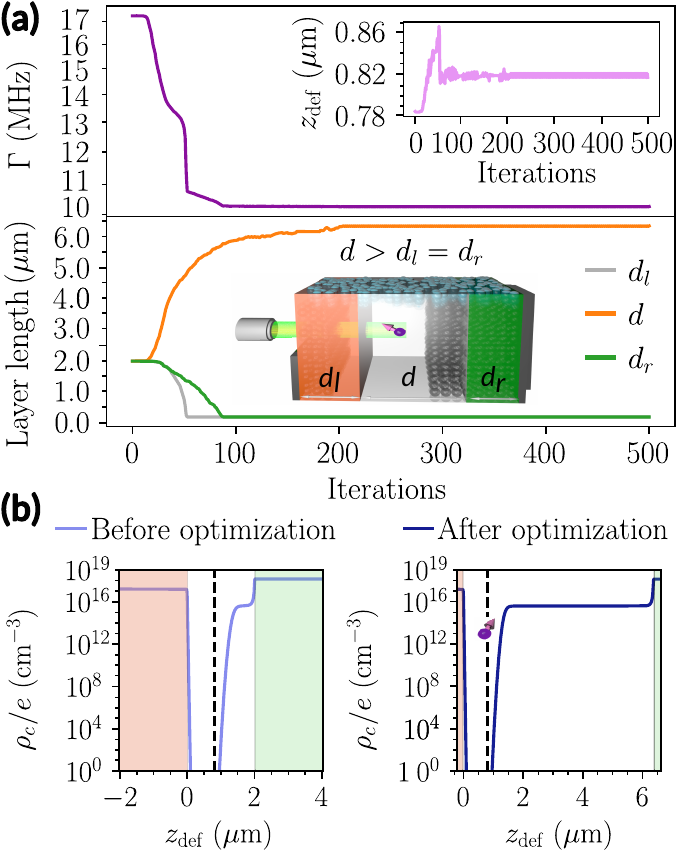}
\caption{Optimization with respect to the lengths of the diode's doping layers. The initial parameters for this simulation are $N_a=$ $7\times10^{18}~\mathrm{cm^{-3}}$, $N_n=$ $4\times10^{15}~\mathrm{cm^{-3}}$, $N_d=$ $1.01\times10^{19}~\mathrm{cm^{-3}}$, $T=300$ $\mathrm{K}$, $V=-15$ $\mathrm{V}$, and $d_l=d_r=d=$ $1$ $\mathrm{\mu m}$. Upper panel: Linewidth reduction and inset showing spin center's position as a function of iteration number. Lower panel: Evolution of $d$, $d_l$, and $d_r$ that allows for minimal optical linewidth $\Gamma$. The size of the lightly-doped $n$-region is considerably bigger than that of the bulk $p$ and $n^+$-regions. This is a consequence of both charge conservation and the application of our electric noise model.}
\label{FIG5}
\end{figure}

\begin{figure*}[t!]
    \includegraphics[width=1\linewidth]{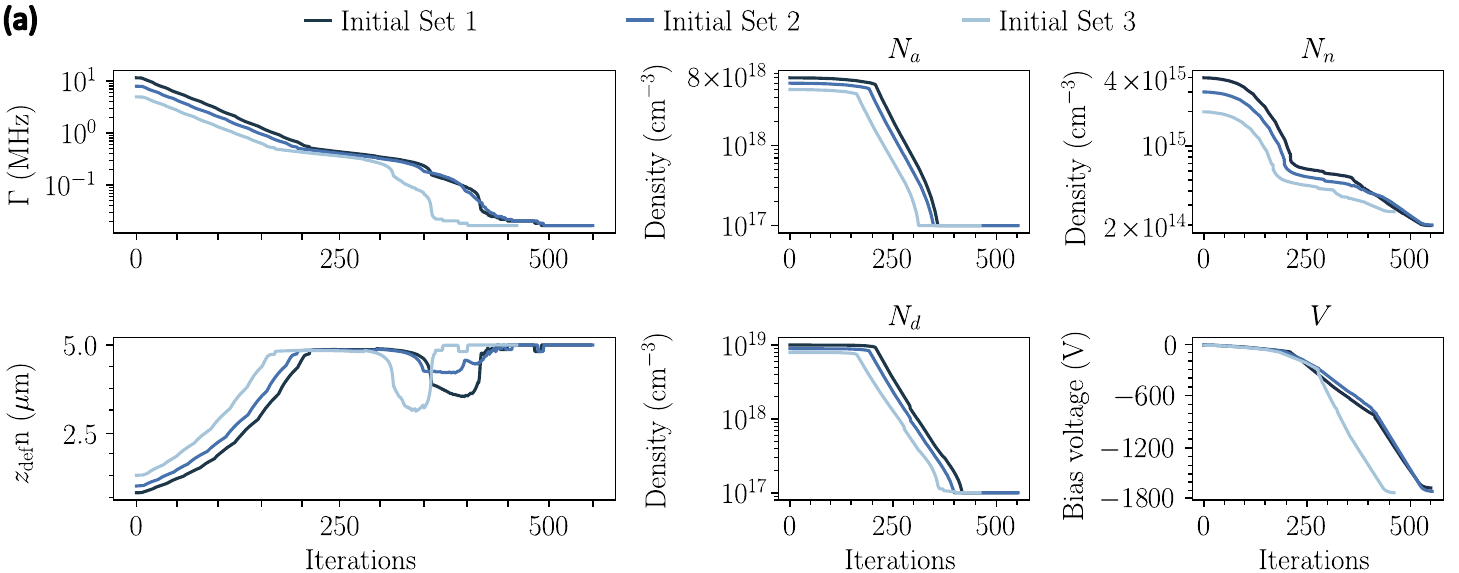}
    \caption{Optimization process for 3 different sets of initial design parameters with displayed iteration-by-iteration behavior of the linewidth $\Gamma$, optimal spin center position and the diode's design parameters $N_a$, $N_n$, $N_d$, and $V$ for $d_l=d_r=$ $0.4$ $\mathrm{\mu m}$ and $d=$ $10$ $\mathrm{\mu m}$. The voltages used for ``Initial Set~1'', ``Initial Set~2'', and ``Initial Set~3'' are $-5$ V, $-6$ V, and $-7$ V, respectively. (b) Same for $d_l=d_r=$ $0.04$ $\mathrm{\mu m}$ and $d=$ $1$ $\mathrm{\mu m}$.  (c) Same for $d_l=d_r=$ $0.004$ $\mathrm{\mu m}$ and $d=$ $0.1$ $\mathrm{\mu m}$.}
    \label{FIG6}
\end{figure*}

\subsection{Optimization with respect to the relative lengths of $p$, $n$ and $n^+$ regions}
\label{RESULTSlen}

Here, we perform the optimization of the linewidth with respect to the lengths of the $p$, $n$, and $n^+$ diode regions, $d_l$, $d$, and $d_r$, respectively, while keeping all other parameters fixed. To explore optimal relative lengths of the diode doping regions, we chose equal initial lengths for all regions, i.e., $d_l=d_r=d=1$~ $\mathrm{\mu m}$. The results are plotted in Fig.~\ref{FIG5} for $N_a=7\times10^{18}$ $\mathrm{cm^{-3}}$, $N_n=4\times10^{15}$ $\mathrm{cm^{-3}}$, $N_d=1.01\times10^{19}$ $\mathrm{cm^{-3}}$, $T=300$ $\mathrm{K}$, $V=-15$ $\mathrm{V}$, and $z_{\text{def}}=0.46$~$\mu$m as the initial spin center position. 

The upper panel of Fig. \ref{FIG5}(a) shows that the optimization of $\Gamma$ leads to a one order of magnitude reduction of the linewidth.
The lower panel of Fig. \ref{FIG5}(a) displays the corresponding change of $d_l$, $d$, and $d_r$ versus iteration, with optimal values achieved by the increase of $d$, while decreasing $d_l$ and $d_r$. Here, increasing $d$ allows for greater separation between the spin center and the fluctuators in the $p$ and $n^+$ regions, whereas diminishing the values of $d_l$ and $d_r$ reduces the number of fluctuators in these regions, as shown by Fig.~\ref{FIG5}(b). The decrease of $d_l$ and $d_r$ ceases at $100$~nm, set as the lower threshold length for $n^+$ and $p$ regions. The inset of Fig.~\ref{FIG5}(a) shows that the optimal spin center position slightly moves away from the $p$-side, and does not reach the middle of the diode, as in Secs. \ref{ResultsVOLT} and \ref{ResultsDOPE}. This happens because across the iterations, the intrinsic region is never fully depleted, as shown by the free-charge profile in Fig. \ref{FIG5}(b). Therefore, $z_{\textrm{def}}$ stagnates to avoid increasing the proximity to these charge carriers.


In short, our results indicate that electric noise reduction in diodes requires the length of the intrinsic $n$-layer ($d$) to be significantly larger than that of the $p$ ($d_l$) and $n^+$-doped ($d_r$) regions \cite{Candido1,nvdiode1,qdots1,qdots4}.

\subsection{Optimization for different diode length regimes with respect to many diode parameters}

\label{ResultsB}

In this section, we optimize the optical linewidth with respect to all doping densities and voltages for various diode total lengths. Fig.~\ref{FIG6} shows the optimization results for a diode with a total length of $10.8$ $\mathrm{\mu m}$ considering three different sets of initial values of $N_a$, $N_n$, $N_d$, and $V$. The initial values of $V$ in ``Initial Set 1'', ``Initial Set 2'', and ``Initial Set 3'' are $-5$ V, $-6$ V, and $-7$ V, respectively; whereas the initial values for the doping densities and spin center position can be read from the panels in Fig.~\ref{FIG6}. The left panels show the behavior of $\Gamma$ and the spin center position as a function of iteration number, whereas the plots on the right side show the corresponding variations of all doping densities and bias voltage. Our optimization with respect to these diode parameters shows an overall two-order-of-magnitude decrease in the linewidth. Here, smaller linewidths are achieved by decreasing doping densities, increasing the magnitude of the bias voltage, and placing the spin defect at the center of the diode. 
Similarly to the results found with respect to a single type of parameters in Secs .~\ref {ResultsVOLT} and \ref{ResultsDOPE}, optimization was obtained via the decreases in the density of fluctuators and increases in their distance from the spin centers. We note that despite the optimal high reverse bias voltages of $-1800$~V, we have not reached the dielectric breakdown regime as $V_{BD}\approx E_{BD}d$ yields $V_{BD}\approx -1900$~V for $d=10.8$~$\mu$m and $E_{BD}=1.9$~MV/m. All doping densities but $N_n$ reach their respective lower thresholds discussed in Sec. \ref{ResultsDOPE}. 

\begin{figure*}[ht]
    \includegraphics[width=1\linewidth]{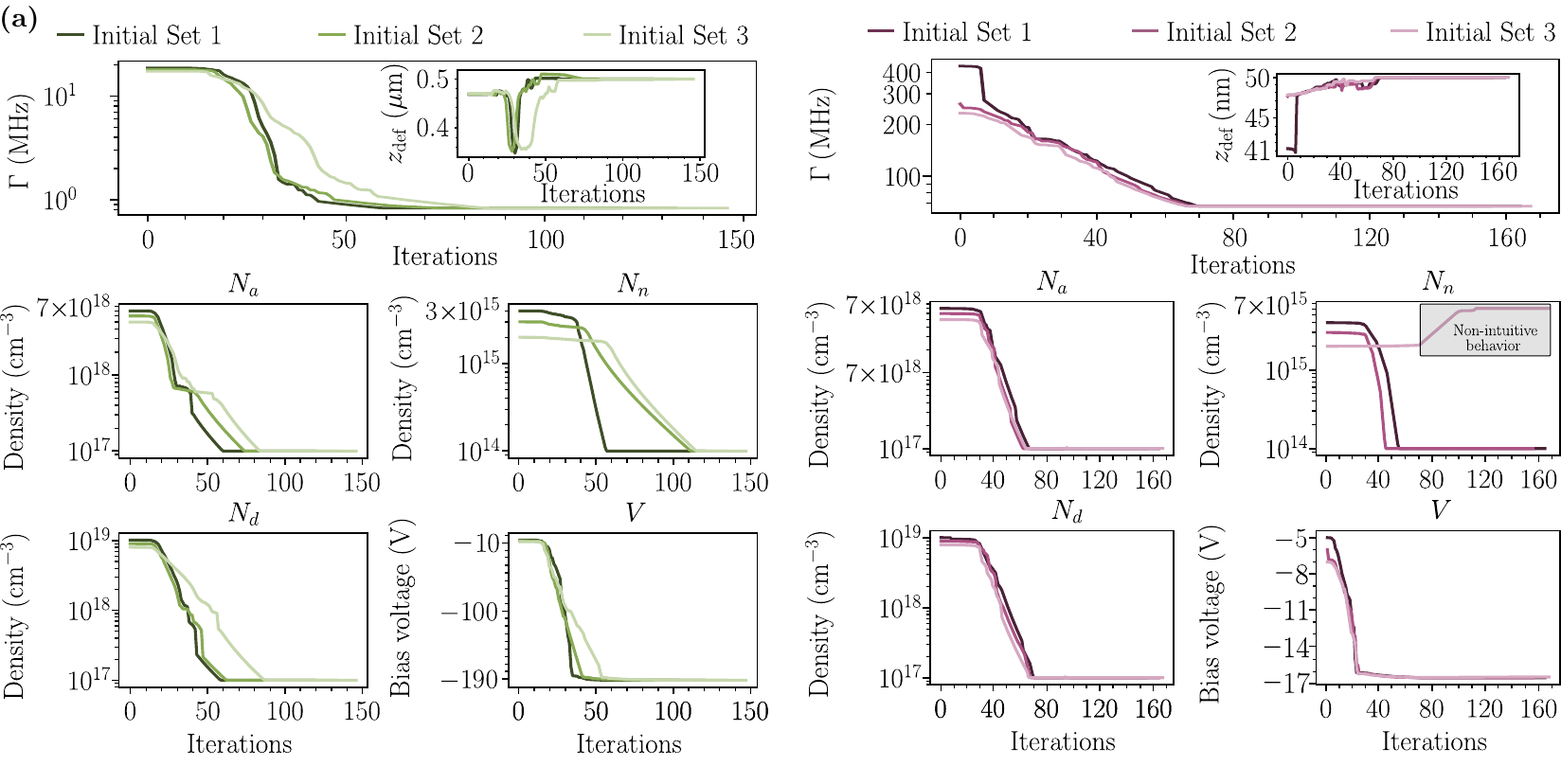}
    \caption{(a) Optimization process for 3 different sets of initial design parameters with displayed iteration-by-iteration behavior of the linewidth $\Gamma$, optimal spin center position and the diode's design parameters $N_a$, $N_n$, $N_d$, and $V$ for $d_l=d_r=$ $0.04$ $\mathrm{\mu m}$ and $d=$ $1$ $\mathrm{\mu m}$. The voltages used for ``Initial Set~1'', ``Initial Set~2'', and ``Initial Set~3'' are $-5$ V, $-6$ V, and $-7$ V, respectively. (b) Same as (a) but for $d=$ $0.1$ $\mathrm{\mu m}$.}
    \label{FIG7}
\end{figure*}

\begin{figure}
    \centering
    \includegraphics[width=1\linewidth]{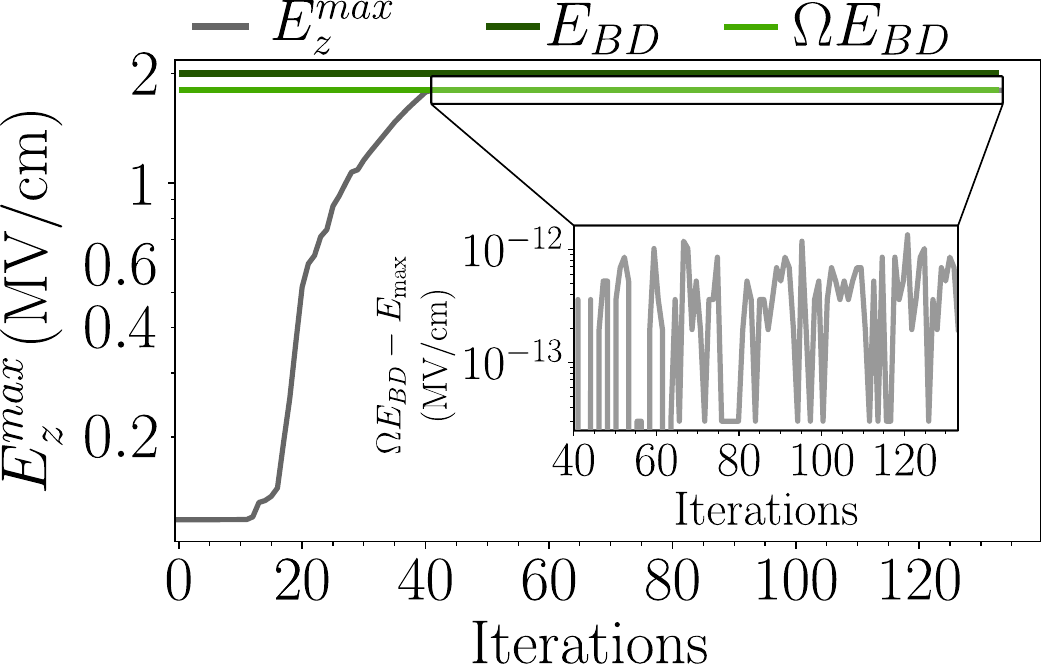}
    \caption{Maximum electric field [$E_{z}^{max}$] inside the diode as a function of the iteration. Horizontal lines represent the electric field breakdown value $E_{BD}$, and a fraction of $\Omega E_{BD}$ ($\Omega<1$) set as the numerical threshold of the electric field. Inset shows $\Omega E_{BD}-E_{z}^{max}$ demonstrating that the maximum value of the electric field never surpasses $\Omega E_{BD}$. }
    \label{figEBD}
\end{figure}

It is well known that different applications of emitters embedded in diodes require different diode lengths. For instance, when coupling emitters to a waveguide or a cavity, small diode lengths ($\lesssim 200$~nm) are required to achieve strong-coupling to single cavity-modes~\cite{qdots4,qdots1}. Conversely, for applications that do not require a cavity, this condition is relaxed, and the total diode lengths can reach $\gtrsim 1$~$\mu$m~\cite{Candido1,nvdiode1,breakd2}. Therefore, we also apply our optimization algorithm for different diode total lengths. Figs.~\ref{FIG7}(a) and (b) show optimization results for fixed total diode lengths of $1.08$ $\mathrm{\mu m}$ and $0.108$ $\mathrm{\mu m}$ with the same initial doping densities and voltages as in Fig.~\ref{FIG6}, and initial spin center positions displayed in their corresponding plots. Here, the overall larger values of the linewidth are a consequence of the smaller diode sizes and hence shorter separations between the spin center and charge fluctuators. Similarly to 
the results in Fig.~\ref{FIG6}, we also observe the decrease of the linewidth due to the decrease of the densities and the increase of the magnitude of the reverse bias voltage, together with optimal placement of the spin center at $z_{\text{def}}\approx d/2$. However, differently from Fig.~\ref{FIG6}, here the magnitudes of the voltage converge to smaller values of $\approx-190$~V for Fig.~\ref{FIG7}(a) and $\approx-16.5$~V for Fig.~\ref{FIG7}(b). 
Using $V_{BD}\approx E_{BD}\,d$ with $E_{BD}\lesssim 2$~MV/cm~\cite{GODIGNON2024108347}, we obtain $V_{BD}\sim-190$~V for $d=1$ $\mu$m, and $V_{BD}\sim-19$~V for $d=0.1$ $\mu$m, thus demonstrating that the further decrease of voltage is not possible due to the dielectric breakdown constraint being respected. {To illustrate this better, in Fig.~\ref{figEBD} we plot the maximum value for the electric field within the diode as a function of iteration number for the case of Fig.~\ref{FIG7}(a). The plot clearly shows that once the maximum electric fields approach the threshold set in our algorithm $\Omega E_{BD}$, but it never crosses the line, demonstrating the avoidance of entering the dielectric breakdown regime.}

Special attention is devoted to the observed increase in $N_n$ for ``Initial Set 3" [Fig. \ref{FIG7}(b)] between iterations 70 and 160, which is seemingly counterintuitive, as noise suppression typically mandates lower doping densities. Physically, this increase facilitates the expansion of the $p$-side depletion width $d_p$ to maintain charge neutrality (i.e., $N_a d_p = N_n \tilde{d}_n$) \cite{Candido1, SEP}. This expansion, acting in tandem with the decreasing values of $N_a$ and $N_d$, and the increasing reverse bias $|V|$, effectively minimizes the total fluctuator count. Nevertheless, this optimal trajectory in parameter space becomes restricted once $N_a$, $N_d$, and $|V|$ reach their respective physical constraints (i.e., lower doping thresholds and the dielectric breakdown limit). In this regime, the full gradient $\nabla_{\textbf{p}} \,\Gamma$ effectively collapses onto its partial derivative with respect to $N_n$. Because following this sole remaining degree of freedom does not yield further improvements in $\Gamma$ during the iteration window, the algorithm invokes a stability mechanism by reducing the learning rate to avoid over stepping in parameter space (see Appendix \ref{AppD}). Consequently, according to Eq. (\ref{gradient}), $N_n$ saturates near iteration $110$. Since at this stage all optimization parameters have saturated, we obtain convergence of the optimization trial shown in Fig.~\ref{FIG7}(b).


In summary, our results indicate that for $\sim10$~$\mu m$-long diodes, optimization is obtained for large reverse bias voltages ($\sim 1800$~V) to maximize the depletion lengths of the doped regions, in addition to small densities ($N_n\sim 10^{14}$~cm$^{-3}$ $N_{a/d}\sim 10^{17}$~cm$^{-3}$) to diminish the charge noise from the non-depleted regions.
For diode lengths of $1$ $\mu$m and $0.1$ $\mu$m, optimizing also involves decreasing densities and increasing reverse voltages, with the distinction that the magnitude of $V$ is now bounded by the corresponding maximum electric field yielding dielectric breakdown.
Finally, for the optimal parameters achieved in this section, the insulating regions are nearly fully depleted, suggesting that the position minimum should occur at the location farthest from the bulk regions, which is approximately at the middle of the diode.

\section{Conclusion}
In summary, our work provides guidance for the manufacture and engineering of diode hosts to maximize coherence of embedded quantum units. In general, our work reveals that promising diodes require large bias voltages, low doping densities, and intrinsic layers larger than the diode's $p$- and $n$-bulk regions, provided that all relevant physical constraints (e.g., density and length thresholds, avoidance of dielectric breakdown) are satisfied. The formalism outlined here can be applied to any quantum emitter embedded in diodes and opens avenues for understanding how to increase the coherence of quantum emitters and solid-state spin qubits.


\begin{acknowledgments}
We thank M. E. Flatté, R. Uppu, V. F. Rodrigues and H. C. Hammer for useful discussions. 

\end{acknowledgments}

\appendix



\section{Quasi-Fermi level $\mu_l$ and boundary value at diode's right end}
\label{AppA}



Here, we outline how to calculate the quasi-Fermi level $\mu_l$ and $\phi\,(z=d_r)=\phi_\infty$ \cite{Candido1}. First, to calculate $\mu_l$ we ensure that charge neutrality is enforced at $z=-d_l$, i.e., $\rho\,(z=-d_l)=0$ [Eq.~(\ref{EQ8_modified})], yielding
\begin{equation}
n_{i}(T,E_{g})e^{\frac{\epsilon_{i}-\mu_l}{k_{B}T}} - n_{i}(T,E_{g})e^{\frac{\mu_l+e V-\epsilon_{i}}{k_{B}T}} - \frac{N_{a}}{2e^{\frac{\epsilon_{a}-\mu_l}{k_{B}T}}+1} = 0,
\label{EQA1}
\end{equation}
\noindent assuming  $\phi(z=-d_l)=0$. Similarly, we find $\phi(z=d_r)=\phi_\infty$ {by applying} charge neutrality at $z=d_r$. Using the value obtained for $\mu_l$ in Eq. (\ref{EQA1}), we solve for $\phi(z=d_r)$ via 
\begin{equation}
\begin{split}
\frac{N_{d}}{2e^{\frac{\mu_l+eV +e\phi(z=d_r)-\epsilon_{d}}{k_{B}T}}+1} 
- n_{i}(T,E_{g})e^{\frac{\mu_l+eV+e\phi(z=d_r)-\epsilon_{i}}{k_{B}T}} \\
+ n_{i}(T,E_{g})e^{\frac{\epsilon_{i}-\mu_l-e\phi(z=d_r)}{k_{B}T}} = 0,
\label{EQA2}
\end{split}
\end{equation}
which yields $\phi(z=d_r)=\phi_\infty$.

\section{Modified Poisson equation}
\label{AppB}

To avoid numerical divergencies and facilitate convergence of our Poisson's equation [Eq.~(\ref{EQ8})], we first define the dimensionless electrostatic potential $\Psi\left(z\right)=[e\phi\left(z\right)+\mu_{l}-\epsilon_i-E_{g}/2]/{k_{B}T}$, and implement the regulating Sigmoid function from Ref. \cite{solver} that incorporates softly bounding values into our total charge density $\rho(z)$ [Eq.~(\ref{EQ8_modified})]. The Sigmoid function is defined as \cite{solver}
\begin{equation}
S\left(\Psi\left(z\right),a\right)=a\ln\left(10\right)\tanh\left(\frac{\Psi\left(z\right)}{a\ln\left(10\right)}\right),
\label{EQB1}
\end{equation}
 where $a=308$ is the upper bound machine-precision for Python, which is our chosen language to perform the computations. Using  \( n_{i}(T, E_{g}) = \sqrt{N_{c}(T)P_{v}(T)}e^{-\frac{E_{g}}{2k_{B}T}} \) and enclosing the arguments of the exponential functions in $\rho(z)$ [Eq.~(\ref{EQ8_modified})] with $S$, we obtain the modified one-dimensional Poisson's equation
\begin{equation}
\frac{\partial^{2}\Psi\left(z\right)}{\partial z^{2}} = -\frac{e^{2}}{k_{B}T}\frac{\rho\left(z,\Psi\left(z\right)\right)}{\epsilon},
\label{EQB8}
\end{equation}
with

\begin{widetext}
\begin{equation}
\begin{aligned}
\rho\!\left(z,\Psi(z)\right)
= {} & e\sqrt{N_c(T)P_v(T)}
\left[
- e^{S\left(\Psi(z)+\frac{V}{k_B T},\,a\right)}
+ e^{S\left(-\Psi(z)-\frac{E_g}{k_B T},\,a\right)}
\right] - \frac{N_a}{2e^{S\left(-\Psi(z)+\frac{\epsilon_a-\epsilon_i-\frac{1}{2}E_g}{k_B T},\,a\right)}+1}
\;\Theta(z+d_l)\,\Theta(-z) \\[6pt]
& + \frac{N_n}{2e^{S\left(\Psi(z)+\frac{\frac{1}{2}E_g-\epsilon_d+\epsilon_i+V}{k_B T},\,a\right)}+1}
\;\Theta(d-z)\,\Theta(z)  + \frac{N_d}{2e^{S\left(\Psi(z)+\frac{\frac{1}{2}E_g-\epsilon_d+\epsilon_i+V}{k_B T},\,a\right)}+1}
\;\Theta(z-d)\,\Theta(d+d_r-z).
\label{mod_rho}
\end{aligned}
\end{equation}
\end{widetext}



%
%

The boundary conditions for $\Psi(z)$ are obtained by evaluating $\phi(z=-d_l)=0$ and $\phi(z=d_r)$ (obtained from the procedure in Appendix \ref{AppA}) into the definition of $\Psi(z)$, i.e.,
\begin{equation}
\Psi\left(z=-d_l\right)=\frac{\mu_l-\epsilon_i-\frac{1}{2}E_{g}}{k_{B}T},
\label{EQB9}
\end{equation}
\begin{equation}
\Psi\left(z=d_r\right)=\frac{e\phi\left(z=d_r\right)+\mu_l-\epsilon_i-\frac{1}{2}E_{g}}{k_{B}T}.
\label{EQB10}
\end{equation}

\section{Linewidth Formalism}
\label{AppLNW}

Here, we use Kubo's lineshape formalism \cite{lineshape1} to obtain both the lineshape and linewidth associated with spin defects embedded in diodes. For a Gaussian process, the variance of the phase arising from any time-dependent fluctuating field $A(t)$ is given by
\begin{align}
\delta \phi_i^2(t)
= C^2 \int_0^t \int_0^t {\Pi_{A_{i}}}(t',t'')\, dt'' dt',
\label{eq_lkg_1}
\end{align}
where $\Pi_{A_{i}}(t',t'')\equiv \langle \delta A_i(t') \, \delta A_i(t'') \rangle \ $ is the two-point time autocorrelation of $A(t)$ and $C$ is a proportionality constant that depends on the magnetic or electric character of the fluctuating field. Assuming temporal translational symmetry of the noise i.e., $\Pi_{A_i}(\tau=t''-t') = \Pi_{A_i}(t'',t')$ and defining the spectral noise density as $S_{A_i}(\omega)\equiv\int_{-\infty}^{\infty}\Pi_{A_i}(\tau)e^{-i\omega\tau}$, Eq. (\ref{eq_lkg_1}) can be recast as

\begin{equation}
    \delta \phi_{i}^2(t) = \frac{C^2}{2\pi} \int_{-\infty}^{\infty} d\omega \,S_{A_i}(\omega) \frac{\sin^2(\omega t/2)}{\omega^2/4}.
    \label{eq_lkg_3}
\end{equation}

\noindent Then, the line-shape $I\,(\omega)$ is given by

\begin{equation}
I(\omega) = \frac{1}{2\pi} \int_{-\infty}^{\infty} dt\, e^{-i\omega t} \mathcal{C}(t),
\label{eq_lkg_4}
\end{equation}

\noindent where the coherence function $\mathcal{C}(t)$ is defined as $\mathcal{C}(t)\equiv \exp{[-\frac{1}{2} \delta \phi_i^2(t)}]$. Finally, the linewidth is determined from the FWHM of $I\,(\omega)$. 
Using this formalism, we now derive the optical linewidth arising from charge and magnetic noise in our diode. 

\subsection{Linewidth due to electric noise from majority carriers}

The random motion of majority carriers from the non-depleted diode regions leads to fluctuations in the energy levels of quantum emitters. The nature of such fluctuations is associated with electric charge noise \cite{nvdiode6,FDT1,FDT2,FDT3}, i.e.,
\begin{equation}
\delta \phi_z^2(t) = \frac{C_E^2}{2\pi} \int_{-\infty}^{\infty} S_{E_z} (\omega) \frac{\sin^2(\omega t/2)}{\omega^2/4} \, d\omega,
\label{eq_lkg_5}
\end{equation}
with $C_E = \frac{2}{\hbar}\mu_z$, $\mu_z$ the dipole moment associated with the optical transitions between the levels of our 4H-SiC di-vacancies. Because charge noise normally obeys a $1/f$ SND dominated by low-frequency components \cite{slowNOISE1}, we assume that it falls within the quasi-static or slow-noise regime. This approximation holds when the correlation time $\tau_c$ of such noise is much greater than the system's coherence time scale~\cite{lineshape2,slowNOISE2,slowNOISE_JUST}, effectively keeping the noise amplitude constant during experimental measurements \cite{lineshape2}. Thus, in this limit, the integrand in Eq. (\ref{eq_lkg_5}) becomes $\frac{\sin^2(\omega t/2)}{\omega^2/4} \approx t^2$, yielding
\begin{equation}
\delta \phi_i^2(t) \approx C_E^2\, \Pi_E\,(\tau=0)\,t^2 = C_E^2\, \delta E_z^2\,t^2= 2\gamma_E^2\,t^2,
\label{eq_lkg_6}
\end{equation}
where $\gamma_E\equiv\frac{\mu_z \delta E_z}{h}$ and $\delta E_z^2\equiv \frac{1}{2\pi}\int_{-\infty}^{\infty} S_E(\omega) \,d\omega\,=\Pi_E\,(\tau=0)$. For this, the resulting line-shape is Gaussian,
\begin{align}
I\,(\omega)
&= \frac{1}{2\pi} \int_{-\infty}^{\infty} dt\, e^{-i\omega t} \exp{[\mathcal{C}(t)]}, \notag\\
&= \frac{1}{2\sqrt{\pi}\gamma_E}\,
\exp\!\left[-\frac{\omega^2}{4\gamma_ E^2}\right],
\label{eq_lkg_7}
\end{align}
with corresponding linewidth
\begin{equation}
    \Gamma=\alpha\sqrt{\Pi_E\,(\tau=0)}\,,
    \label{eq_lkg_8}
\end{equation}
where $\alpha\equiv \sqrt{\frac{2 \ln 2}{\pi}}\,\frac{\mu_z}{ \hbar}$. In Appendix \ref{AppC1}, we outline the model used to characterize $|\delta \mathbf{E}|^2$ \cite{Candido1}, allowing us to understand the linewidth due to the non-depleted regions of the diode.

\subsection{Linewidth due to electric noise from carriers in the depletion region}
\label{AppLWelecSURF}

\textcolor{black}{Here,} we consider the \textcolor{black}{noise generated by carriers generated within the diode's depletion region \cite{LEAKS8,LEAKS5,GTAT} and derive its resulting linewidth, which is associated with the leakage current in such region \cite{leak2,leak4}. Since leakage current is mostly dominated by carrier generation and trapping at surface states \cite{LEAKS5,LEAKS7,SURFACE_LEAK,SURFACE_LEAK2}, we focus on the linewidth due to surface charge noise.} For a spin center at a depth $x_{\text{def}}$, we use the model from Ref. \cite{rates}, which defines the SND generated by a fluctuating dipole surface density as

\begin{align}
S_{E'_z}^{2D}(\omega,x_{\text{def}},x') &=
\left( \frac{e}{4\pi\epsilon} \right)^2
\frac{\pi\, n_S(x') (3 + \cos 2\theta)}
{8 |x_{\text{def}}-x'|^4}
S_d(\omega),
\label{eq_lkg_9a}
\\[6pt]
S_d(\omega) &\equiv
\frac{d^2 \Gamma_d}{\omega^2 + (\Gamma_d/2)^2}.
\label{eq_lkg_9b}
\end{align}
In Eqs. (\ref{eq_lkg_9a}) and (\ref{eq_lkg_9b}), $S_d(\omega)$ is the Lorentzian spectrum for an electric dipole with variance $d^2$, orientation angle $\theta$ with respect to $z$, and rate $\Gamma_d$. $n_S(x')$ represents the surface fluctuator density. To account for the distribution of defects throughout the volume, we assume a stack of two-dimensional densities $n_S (x')$ separated by the crystal's lattice constant $\delta$. The total noise experienced by the spin center is the sum of the contributions from each discrete plane located at $x'_n=n\,\delta$ with $n=0,1,2,...$, resulting in
\begin{equation}
    S_{E'_z}(\omega, x_{\text{def}}) = \sum_{n} S_{E'_z}^{2D}(\omega, x_{\text{def}}, x_n),
    \label{eq_sum}
\end{equation}
where we have assumed no correlation between different layers. The surface density $n_S(x_n)$ is related to the volume density $n_{V}(x_n)$ near the surface by $n_S(x_n) = n_{V}(x_n)\, \delta$. We then substitute the sum as an integral assuming $\delta\rightarrow0$, yielding
\begin{equation}
    \sum_{n} ( \dots) n_{V}(x_n) \delta \approx \int (\dots) n_{V}(x') dx'.
    \label{eq_Riemann_sum}
\end{equation}
and
\begin{equation}
   S_{E'_z}(\omega, x_{\text{def}}) = C_{E}^2\,  S_d(\omega) a(x_{\text{def}}), 
   \label{eq_lkg_10}
\end{equation}
with $C_E=\frac{2}{\hbar}\mu_z$ and 
\begin{equation}
a(x_{\text{def}}) \equiv \frac{  \pi  (3 + \cos 2\theta)}{8} \left( \frac{e}{4\pi\epsilon} \right)^2 \int_{0}^{D}\frac{n_V(x')}{|x_{\text{def}}-x'|^4}\,dx'.    
\label{eq_lkg_11}
\end{equation}
Then, $\delta\phi^2$ becomes
\begin{equation}
\delta \phi_z^2(t,x_{\text{def}}) = \frac{C_{E}^2\,a(x_{\text{def}})}{2\pi} \int_{-\infty}^\infty S_d(\omega) \frac{\sin^2(\omega t/2)}{\omega^2/4} \, d\omega.
\label{eq_lkg_12}
\end{equation}
Similarly to the majority carrier case, surface charge noise is also slow-noise and as a result  \begin{equation}
\delta \phi_z^2(t,x_{\text{def}}) \approx \frac{C_{E}^2\,a(x_{\text{def}})}{2\pi}\, t^2 \int_{-\infty}^\infty S_d(\omega) \, d\omega.
\label{eq_lkg_13}
\end{equation}
The integration with respect to $\omega$ gives
\begin{align}
 \int_{-\infty}^{\infty} S_d(\omega) \, d\omega 
&= \int_{-\infty}^{\infty} \frac{d^2 \Gamma_d}{\omega^2 + (\Gamma_d/2)^2} \, d\omega \notag\\
&= 2\pi d^2.
\label{eq_lkg_14}
\end{align}
By defining  $\Pi_{E}'(\tau=0,x_{\text{def}})\equiv d^2\,a(x_{\text{def}})$, we obtain
\begin{align}
\delta \phi^2(t,x_{\text{def}}) &\approx C_{E}^2\,\Pi_{E}'(\tau=0,x)\,t^2,\\
C(t,x_{\text{def}}) &= \exp\left[ -\gamma_{E}'^2(x_{\text{def}})\, t^2 \right],
\label{eq_lkg_15}
\end{align}
where $\gamma_E '\,(x_{\text{def}}) \equiv \frac{1}{2}C_{E}^2\,\Pi_{E}'(\tau=0,x)$. Then, the line-shape is given by
\begin{equation}
I(\omega,x_{\text{def}})  = \frac{1}{2\sqrt{\pi}\,\gamma_E'\,(x_{\text{def}})} \exp\left[-\frac{\omega^2}{4 \gamma_E'^2\,(x_{\text{def}})}\right],
\label{eq_lkg_16}
\end{equation}
 which has the same form as Eq. (\ref{eq_lkg_7}) with corresponding linewidth
\begin{equation}
    \Gamma_{E}'\,(x_{\text{def}}) = \alpha\sqrt{\Pi_{E}'(\tau=0,x_{\text{def}})}\,,
    \label{eq_lkg_17}
\end{equation}
with $\alpha\equiv \sqrt{\frac{2 \ln 2}{\pi}}\,\frac{\mu_z}{ \hbar}$.

\subsection{Linewidth due to magnetic noise from carriers in the depletion region}
\label{AppLWmagSURF}

Here, we examine the broadening of the optical linewidth due to magnetic noise arising from the leakage current and modeled as due to G-R carrier transport within the depletion region \cite{LEAKS5,LEAKS8,GTAT}. Specifically, we focus on the longitudinal magnetic field fluctuations of the Zeeman energy splitting. Different from Appendix \ref{AppLWelecSURF}, here the SND is characterized by Johnson-Nyquist noise \cite{JohnsonNyquist1,JohnsonNyquist2}. Because the fluctuation dynamics of Johnson-Nyquist noise are governed by carrier scattering events, occurring on timescales much smaller (i.e., $\tau_c \sim 10^{-13}$~s) \cite{JNnoise} than the typical emitter coherence time (i.e., $T_2^* \sim$ ns - $\mu$s) \cite{JNnoise2}, the noise resides in the motional narrowing or fast-noise limit ($\tau_c \ll T_2^*$) \cite{lineshape2}. Consequently, the magnetic field correlation function can be approximated as a delta function, $\Pi_B (t) \propto \delta(t)$, which results in a Lorentzian line-shape \cite{lineshape1}. In this limit, we can use $\frac{\sin^2(\omega t / 2)}{\omega^2/4} \rightarrow 2\pi \delta(\omega)\,t$, leading to the phase variance
\begin{align}
\delta \phi_z^2(t,x_{\text{def}}) 
&= C_B^2 \, t \int_0^\infty S_{B_z}(\omega,x_{\text{def}}) \, \delta(\omega) \, d\omega, \notag\\
&= C_B^2 \, S_{B_z}(\omega=0,x_{\text{def}}) \, t \,,
\label{eq_lkg_18}
\end{align}
with $C_B= \frac{2 \mu_B \,g}{\hbar} $, $\mu_B$ the Bohr magneton and $g$ is the effective gyro-magnetic ratio associated with the optical transition between the ground and excited states of our spin defects \cite{nvdiode1,nvdiode6}. Similar to what we did for the electric noise in Appendix \ref{AppLWelecSURF}, we assume that the three-dimensional SND can be constructed by summing the noise contributions of two-dimensional layers separated by the crystal's lattice constant $\delta$, i.e., 
\begin{equation}
    S_{B_z}(\omega, x_{\text{def}})\approx\sum_{n} S_{B_z}^{\text{2D}}(\omega, x_{\text{def}}, x_n),
    \label{sum_magn}
\end{equation}
with $x_n=n\,\delta$ and $n=0,1,2, ...$. For $S_{B_z}^{\text{2D}}$, we use the model developed in Ref.~\cite{NOISElkg1}, namely
\begin{equation}
\begin{aligned}
S_{B_z}^{\text{2D}}(\omega,x_{\text{def}},x') 
&= \frac{k_B T \mu_0^2}{16 \pi  |x_{\text{def}}-x'|^2} 
  \int_0^\infty r \, e^{-r} \,\times \\
&\quad\text{Re} \Big[
    \sigma_S^{xx}\Big(\tfrac{r}{2|x_{\text{def}}-x'|}, \omega\Big) + \\
&\quad\sigma_S^{yy}\Big(\tfrac{r}{2|x_{\text{def}}-x'|}, \omega\Big)
  \Big] dr.
\label{eq_lkg_21}
\end{aligned}
\end{equation}
where $\sigma_S^{ii}$ is the 2D conductivity tensor. {The mapping between the 2D and 3D conductivity is given by $\sigma_S^{ii} = \sigma_V^{ii}\,\delta$. Thus, as $\delta\rightarrow 0$, we apply the rule in Eq. (\ref{eq_Riemann_sum}) and recast Eq. (\ref{sum_magn}) as a double integral:}

\begin{equation}
\begin{aligned}
S_{B_z}(\omega,x_{\text{def}},x') 
&\approx \frac{k_B T \mu_0^2}{16 \pi  |x_{\text{def}}-x'|^2} 
 \int_0^D  \int_0^\infty r \, e^{-r} \,\times \\
&\quad\text{Re} \Big[
    \sigma_V^{xx}\Big(\tfrac{r}{2|x_{\text{def}}-x'|}, \omega,x'\Big) + \\
&\quad\sigma_V^{yy}\Big(\tfrac{r}{2|x_{\text{def}}-x'|}, \omega, x'\Big)
  \Big] dr\,dx'.
\label{eq_lkg_19}
\end{aligned}
\end{equation}

Here, we assume $\sigma_V^{ii}$ is given by the frequency-dependent Drude conductivity~\cite{Drude}, i.e., 
\begin{equation}
\sigma_V^{ij}(\omega,x') = \delta^{ij}\frac{e^{2}\,n_V(x')\left(\frac{1}{\tau_{e}} - i\omega\right)}{m_{c}\left(\frac{1}{\tau_{e}^{2}} + \omega^{2}\right)},
\label{EQ27}
\end{equation}
where $\tau_e$ is the electron's scattering time, and $n_V(x')$ is the effective volume density of charges generating leakage current. Then, the coherence function becomes
\begin{equation}
C(t,x_{\text{def}}) = \exp\left[ -\frac{\delta \phi^2(t,x_{\text{def}})}{2} \right] = \exp\left[ -\gamma_B'(x_{\text{def}})\, t \right],
\label{eq_lkg_22}
\end{equation}
with $\gamma_B'(x_{\text{def}})\,=\frac{1}{2}C_B^2\,S_{B_z}(\omega=0,x_{\text{def}})$. The line-shape $I(\omega)$ is then given by
\begin{equation}
    I(\omega,x_{\text{def}}) = 
    \frac{1}{\pi} \frac{\gamma_B'(x_{\text{def}})\,}{\gamma'_B(x_{\text{def}})\,^2 + \omega^2},
    \label{eq_lkg_23}
\end{equation}
which is a Lorentzian distribution centered at zero frequency with linewidth given by its FWHM
\begin{equation}
    \Gamma_B'\,(x_{\text{def}}) = \eta\, S_{B_z}(\omega=0,x_{\text{def}}),
    \label{eq_lkg_24}
\end{equation}
and $\eta\equiv {8 \pi\,\mu_B^2 \,g^2\,}/{h^2}$.

\section{Leakage Current}
\label{AppLKG}

\begin{figure}[h]
\includegraphics[width=0.48\textwidth]{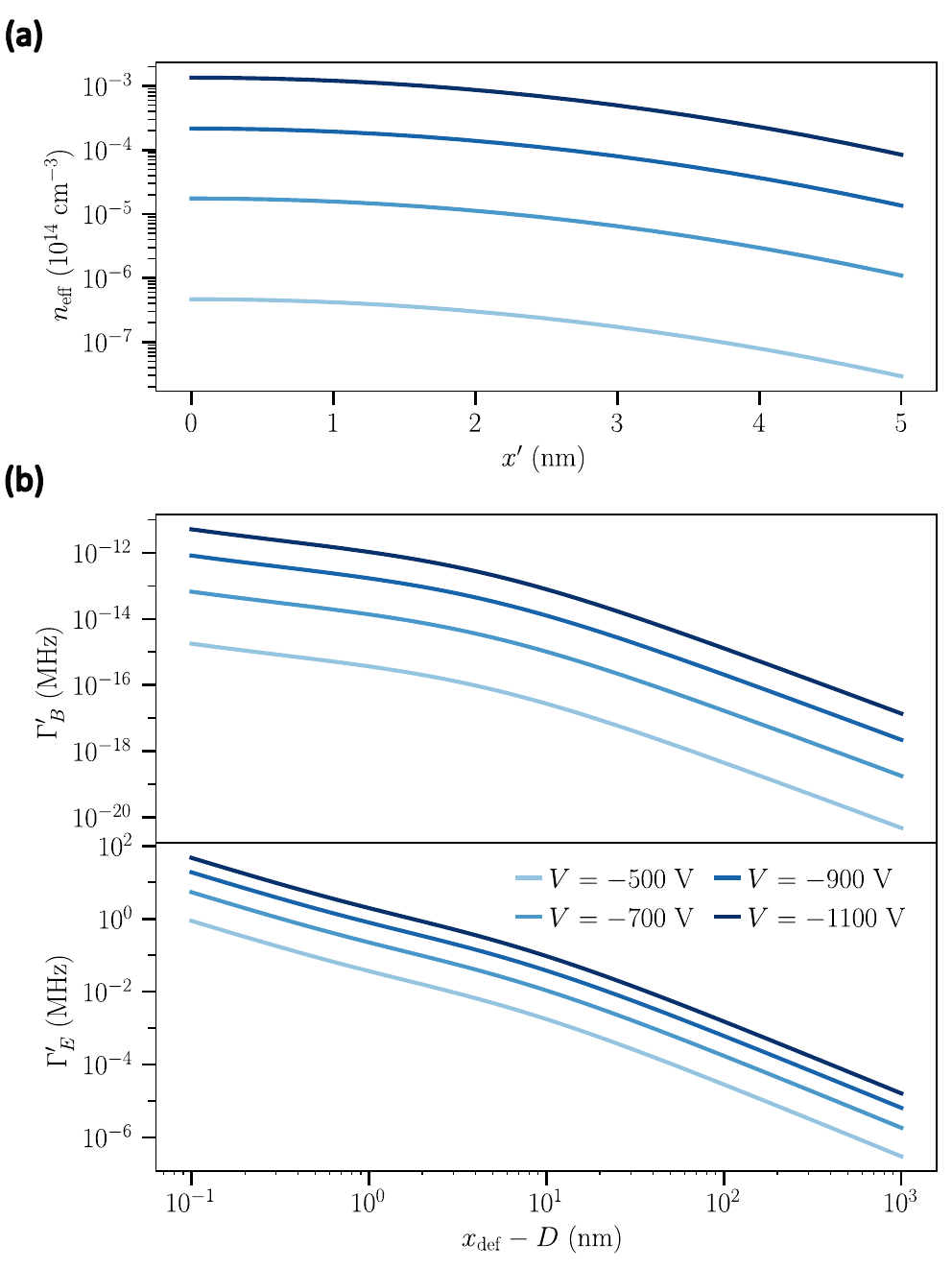}
\caption{
Cause and effect of leakage current present in a 4H-SiC $pnn^+$ diode. (a) Effective density $n_\text{eff}$ produced by leakage current at thermal equilibrium as a function of depth from the diode's surface for reverse bias voltages ranging from $-500$ to $-1100$ $\mathrm{V}$. The depth-dependence of the densities is modeled as a half-Gaussian profile, where its maximum value occurs at the surface of the diode, and it decays as a function of depth inside the diode. This is pictorially evidenced in Fig.~\ref{FIG1}(a). (b) Optical linewidth calculated from electric and magnetic SND as a function of the spin center's separation from the volume containing fluctuators. The linewidths are calculated for the same reverse bias voltages and effective densities shown in (a). This is the effect generated by the leakage current present near the surface of the diode. For this simulation, our initial diode parameters are $N_a=7\times10^{18}$ $\mathrm{cm^{-3}}$, $N_n=$$4\times10^{15}$ $\mathrm{cm^{-3}}$, $N_d=$$1.01\times10^{19}$ $\mathrm{cm^{-3}}$, $T=300$ $\mathrm{K}$, $d_l=d_r=$ $0.4$ $\mathrm{\mu m}$, $d=$ $10$ $\mathrm{\mu m}$, $D=\text{FWHM}=$ $5$ $\mathrm{nm}$, and $\delta=0.1$ nm.}
\label{FIG9}
\end{figure}


 \


 Modeling the leakage current theoretically is not a trivial task, as multiple mechanisms can give rise to it, including drift-diffusion \cite{LEAKS5,metalSEMI1}, electron-hole pair generation and recombination \cite{LEAKS8,LEAKS9}, and carrier tunneling \cite{LEAKS10,LEAKS11,LEAKS12}. While the textbook drift-diffusion model \cite{LEAKS5,metalSEMI1} calculates the absolute theoretical minimum leakage caused by pure thermal excitation, its contribution scales with the square of the intrinsic carrier concentration $n_i^2\propto e^{-E_g/2k_B T}$ \cite{LEAKS5,metalSEMI1} and is therefore vanishingly small in wide band-gap materials, such as 4H-SiC $p$-$n$ diodes. For these materials, the reverse leakage current is primarily driven by field-enhanced \cite{thermal1,HURKX,field_enhancement} carrier generation \cite{LEAKS8}, and trap-assisted tunneling (TAT) \cite{field_enhancement,tat1,tat2} of localized defects. These mechanisms are responsible for the electronic transport within the diode's depletion region \cite{LEAKS8,LEAKS5,GTAT}, where intermediate trap states and phonons act as ``stepping stones'' for carriers traversing within the wide band-gap \cite{field_enhancement}. This process is mediated by the Poole-Frenkel (PF) effect \cite{PF3,PF1,PF2}, in which local electric fields lower the potential barriers of localized states, thereby enhancing carrier emission. To accurately model this behavior, we employ the Hurkx formalism \cite{HURKX}, which incorporates a field-dependent enhancement factor into the standard Shockley-Read-Hall (SRH) generation-recombination framework \cite{LEAKS8}. Although initially developed for silicon, Hurkx's theory \cite{HURKX} provides a unified analytical treatment for the field-dependent emission and tunneling processes observed in SiC $p\text{-}i\text{-}n$ diodes \cite{field_enhancement,tat1,tat2}.


In Eqs. (\ref{eq:lwdth_elec}) and (\ref{eq:lwdth_mag}), the linewidth associated with the leakage current flowing across the depletion region is generally dependent on the density of surface defects \cite{SpinCenter1,rates,SURFACE_LEAK,SURFACE_LEAK2},  suggesting that the leakage current $J$ originates due to the motion of carriers near the surface. Thus, a method for linking the experimentally measured leakage current to an effective carrier density is needed. In the section below, we present our formalism used to calculate the steady-state reverse leakage current via Hurkx's model \cite{HURKX}, as well as its associated effective carrier density, which is used to compute the SND related to the reverse leakage current in our diodes.

\subsection{Derivation of $J$ and $n_{\text{eff}}$}

Here, we derive the reverse leakage current and its associated effective density. We consider a diode's volume containing a density $N_t$ of deep-level traps initially located at energy level $\epsilon_{t,0}$ within the band-gap. As discussed in the introduction to this Appendix, the trap energy level is modified via the PF effect, such that $\epsilon_t=\epsilon_{t,0} - \beta_{PF} \sqrt{E\,(z)}$ \cite{PF3,PF1,PF2}, where $E\,(z)$ is the electric field inside the diode and $\beta_{PF} = \sqrt{q^3 / (\pi \epsilon)}$, with $q$ as the fundamental charge and $\epsilon$ as the permitivity constant of our crystal.
Let $n_t(z,t)$ be the density of traps occupied by electrons at the diode's position $z$ and time $t$. The density of empty traps is therefore $N_t - n_t(z,t)$. Our goal is to find $n_t(z,t)$ when subjected to fluctuations in the occupation of traps. The rate of change in the density of the occupied traps $\frac{dn_t}{dt}$ is determined by four competing processes \cite{LEAKS8,tat2}: electron capture, electron emission, hole capture, and hole emission. The governing differential equation is
\begin{equation}
\frac{dn_t}{dt} = \underbrace{c_n n (N_t - n_t)}_{\text{Capture } e^-} - \underbrace{e_n n_t}_{\text{Emit } e^-} - \underbrace{c_p p n_t}_{\text{Capture } h^+} + \underbrace{e_p (N_t - n_t)}_{\text{Emit } h^+}.
\label{eq:full_rate}
\end{equation}
Here, $c_{n(p)} = v_{\text{th}} \sigma_{n(p)}$ and $e_{n(p)}$ represent the capture and emission coefficients for electrons and holes, respectively, with thermal velocity given by $v_{\text{th}} = \sqrt{3k_B T/m^*}$, where $m^* = 0.37m_0$ is the effective electron mass for 4H-SiC. The emission coefficients are defined as $e_n = c_n N_c\,(T) \exp[-(\epsilon_c - \epsilon_t)/k_B T]$ and ${e_p = c_p N_v\,(T) \exp[-(\epsilon_t - \epsilon_v)/k_B T]}$. $n$ and $p$ denote, respectively, the free electron and hole carrier concentrations within the depletion region at thermal equilibrium \cite{LEAKS8} (obtained by solving Eq. \ref{EQ8}), while $\sigma_{n(p)}$ are the capture cross sections.  In the depletion region of a reverse-biased diode, the free carrier concentrations $n$ and $p$ are negligible compared to their values within the diode's non-depleted region, so we assume $n = p \approx 0 $, yielding
\begin{equation}
\frac{dn_t}{dt} = e_p(N_t - n_t) - e_n n_t.
\label{eq:simple_rate}
\end{equation}
For a position-dependent high electric field $E\,(z)$, an electron bound to a defect need not jump the defect's full energy barrier (i.e., $\epsilon_c - \epsilon_t$). It can also tunnel through the potential barrier into the conduction band \cite{HURKX,tat1,tat2}.
Following Hurkx's model \cite{HURKX}, the thermal emission rate is enhanced by a factor $\Gamma_{n,p}$ to account for tunneling effects:
\begin{equation}
\Gamma_{n,p}[E\,(z)] = \frac{\epsilon_c - \epsilon_v}{k_B T} \int_{0}^{1} \exp\left( \frac{\epsilon_c - \epsilon_v}{k_B T} u - K u^{3/2} \right) du.
\end{equation}
where $u$ a dimensionless integration variable representing normalized energy, and $K \equiv \frac{4}{3} \frac{\sqrt{2m^*} (\epsilon_c - \epsilon_v)^{3/2}}{e \hbar |E\,(z)|}$ the tunneling characteristic parameter. This allow us to define the field and position-dependent emission rates
\begin{equation}
e_n[E\,(z)] = e_{n,th} [1 + \Gamma_n(E)],
\label{em_rate1}
\end{equation}
\begin{equation}
e_p[E\,(z)] = e_{p,th} [1 + \Gamma_p(E)],
\label{em_rate2}
\end{equation}
where $e_{n,th} = c_n N_c(T) \exp\left( -\frac{\epsilon_c - \epsilon_t}{kT} \right)$ and $e_{p,th} = c_p N_v(T) \exp\left( -\frac{\epsilon_t-\epsilon_v}{kT} \right)$. Eq. (\ref{eq:simple_rate}) can be solved analytically, yielding
\begin{equation}
n_t(z,t) = \frac{e_p N_t}{e_n + e_p} + \left[ n_t(z,0) - \frac{e_p N_t}{e_n + e_p} \right] \exp\left[ -(e_n + e_p)t \right],
\label{density_solution}
\end{equation}
with $e_n$ ($e_p$) given by Eq. (\ref{em_rate1}) [Eq. (\ref{em_rate2})]. In the limit of $t\rightarrow\infty$, we obtain the steady-state solution
\begin{equation}
    n_t^{ss}(z) \equiv  n_t^{ss}(z,t\rightarrow\infty)= N_t \frac{1/e_n}{1/e_n + 1/e_p}.
    \label{ss_solution}
\end{equation}

To determine the leakage current $J$, we need to find the generation rate $G_\text{TAT}$ \cite{GTAT}, which quantifies how many carriers per unit volume per unit time escape the trap centers and reach the conduction band. Such a rate is found by setting the left-hand side of Eq. (\ref{eq:simple_rate}) equal to zero, substituting Eq. (\ref{ss_solution}) into either term, and identifying such a combination with a generation with the generation rate, i.e.,

\begin{equation}
\begin{split}
G_\text{TAT}(z) &= e_p(N_t - n_t) = e_n n_t \\
&= N_t \frac{e_{n,th}(1+\Gamma_n)\, e_{p,th}(1+\Gamma_p)}
{e_{n,th}(1+\Gamma_n) + e_{p,th}(1+\Gamma_p)} .
\end{split}
\label{gen_rate}
\end{equation}

\noindent Then, the leakage current measured at the right end of the depletion region is given by \cite{GTAT}

\begin{equation}
     J_{TAT} = q \int_{-d_p}^{\tilde{d}_n +d_n} G_{TAT}(z) \, dz ,
     \label{lkg_curr_right_end}
\end{equation}
where $\tilde{d}_n +d_n$ is the length of the depleted intrinsic plus $n^+$ diode regions, and $d_p$ is the length of the depleted $p$ diode region. To find the carrier density associated with the current density in Eq. (\ref{lkg_curr_right_end}), we assume that the transport mechanism is dominated by drift, due to the high electric fields in the diode's depletion region \cite{SEP,metalSEMI1,LEAKS5}. As a result, $J_n(z) = q n(z) v_d(z)$ at equilibrium, where $J_n(z)$, $n(z)$, and $v_d(z)$ are the spatially and field-dependent leakage current, effective carrier density $n_{\text{eff}}$, and drift velocity, respectively. Because the carrier concentration, current density, and drift velocity are spatially dependent across the depletion region, we assume that the effective density $n_{\text{eff}}$ is given by the average across the entire region. This approximation is necessary because carriers do not appear instantaneously at the depletion boundaries; rather, the observed carrier population at any point $z$ is the integrated result of generation and transport events occurring throughout the depletion volume. Thus, a local value at the boundary cannot accurately represent the ensemble carrier-trap interaction governing the TAT process. The average effective density quantifies the total number of charge fluctuators within the diode's depletion region, which is the input to Eqs. (\ref{eq:lwdth_elec}) and (\ref{eq:lwdth_mag}). As a result, the effective density associated with the experimentally-measured leakage current in Eq. (\ref{lkg_curr_right_end}) is

\begin{equation}
n_{\text{eff}} = \frac{1}{W} \int_{-d_p}^{\tilde{d}_n +d_n} \frac{1}{v_d(z)} \left[ \int_{-d_p}^{z} G_{TAT}(z') \, dz' \right] \, dz.
\label{effective_dens}
\end{equation}

To determine $v_d(z)$ for carriers through the high-field depletion region, we employ the standard field-dependent mobility model for 4H-SiC described by Roschke and Schwierz \cite{vdrift2}, which accounts for velocity saturation at high electric fields, ensuring a realistic estimate of the carrier density $n_{\text{eff}}$ derived from the leakage current. The drift velocity is defined as $v_d\,(z) = \mu[E(z)]E(z)$, with the field-dependent mobility given by \cite{v_drift,vdrift2}
\begin{equation}
\mu[E(z)] = \frac{\mu_{0}}{\left[1 + \left(\frac{\mu_{0}E}{v_{\text{sat}}}\right)^\beta\right]^{1/\beta}},
\label{EQ18}
\end{equation}
where $\mu_{0}$ is the low-field mobility, $v_{\text{sat}}$ is the saturation velocity, and $\beta$ is a parameter used to describe the abruptness of the transition between linear and saturated regimes for $v_d$. Both $\beta$ and $v_{\text{sat}}$ are temperature and material-dependent fitting parameters \cite{vdrift2}. The low-field mobility depends on the total doping concentration ($N_{a}+N_{n}+N_{d}$) via~\cite{v_drift,vdrift2}
\begin{equation}
    \mu_0 = \mu_{\text{min}} + \frac{\mu_{\text{max}} - \mu_{\text{min}}}{1 + \left(\frac{N_{a} + N_{n} + N_{d}}{N_{\text{ref}}}\right)^\alpha}.
    \label{EQ_LowField} 
\end{equation}

For electrons travelling perpendicular to the c-axis in 4H-SiC at 300 K, we utilize the fitting parameters determined in Ref. \cite{vdrift2}: $\mu_{\text{max}} = 950$ cm$^2$/Vs, $\mu_{\text{min}} = 40$ cm$^2$/Vs, $N_{\text{ref}} = 2 \times 10^{17}$ cm$^{-3}$, $\alpha = 0.76$, and a saturation velocity of $v_{\text{sat}} = 2.4 \times 10^7$ cm/s with $\beta = 0.85$. Note that in the depletion region of a reverse-biased diode, the electric field $E$ typically far surpasses the saturation field ($E \gg v_{\text{sat}}/\mu_0$). Consequently, the carriers operate deep within the velocity saturation regime (i.e., $v_d \approx v_{\text{sat}}$). Despite this, $n_{\text{eff}}$ in Eq. (\ref{effective_dens}) considers the fact that $v_d$ varies spatially for completeness. Additionally, because the diode is at thermal equilibrium, $J_{\text{electron}}(z)=J_{\text{hole}}(z)$. Thus, $J\,(z=\tilde{d}_n + d_n)= J_{\text{electron}}$ \textcolor{black}{, since $J_{\text{hole}}\,(z=\tilde{d}_n + d_n)=0$ \cite{SEP,metalSEMI1,LEAKS5}}. \


The effective density of impurities $n_{\text{eff}}$ is assumed to decay exponentially with distance from the diode's surface up to some depth $D$, as displayed in Fig. \ref{FIG1}(a). In our work, we consider a half-Gaussian spatial decay with a FWHM $= 5$ nm (similar to Ref. \cite{LEAKS7}), and peak value given by $\max_{x'}[n_V\,(x')]=n_V\,(0)=n_{\text{eff}}$. As a result, the depth-dependent fluctuator density associated with reverse leakage current is
\begin{equation}
n_V\,(x')=n_{\text{eff}}\,f(x'),
\label{eq:depth_dens_gauss}
\end{equation}
where $f(x')$ is our half-Gaussian decaying distribution. To ensure that the total number of fluctuators in our model corresponds exactly to the number of carriers involved in the conduction process, we enforce that $\int_0^D n_{V}(x') dx' = n_{\text{eff}} D$. Fig. \ref{FIG9}(a) shows $n_V\,(x')$ for large reverse bias voltages. Such a depth-dependent effective density is our input needed to calculate the electromagnetic noise [i.e., Eqs. (\ref{eq_lkg_10}), (\ref{eq_lkg_11}), (\ref{eq_lkg_19}), and (\ref{EQ27})] and the resulting linewidth due to the leakage current given by Eqs. (\ref{eq_lkg_17}) and (\ref{eq_lkg_24}). 

To ensure consistency with experimental results, we calculate $J$ via Eq. (\ref{lkg_curr_right_end}) for the diodes studied in Ref. \cite{nvdiode1}. The inset in Fig.~\ref{FIG9}(a) \textcolor{black}{[TO BE INCLUDED]} displays our predicted reverse leakage current for the diode design parameters from Ref. \cite{nvdiode1}, showing good agreement between the experiment and theoretical model. The diode's initial parameters involved in this simulation are $N_a=7\times10^{18}$ $\mathrm{cm^{-3}}$, $N_n=4\times10^{15}$ $\mathrm{cm^{-3}}$, $N_d=1.01\times10^{19}$ $\mathrm{cm^{-3}}$, $N_t=10^{17}$ $\mathrm{cm^{-3}}$, $\sigma_n=\sigma_p=10^{-15}$ $\mathrm{cm^{2}}$, $T=300$ $\mathrm{K}$, $d_l=d_r=$ $0.4$ $\mathrm{\mu m}$, $d=$ $10$ $\mathrm{\mu m}$, $\delta=0.1$ nm, and $D=\text{FWHM}=5$ $\mathrm{nm}$. Here, $\delta$ represents the lattice constant of our crystal.

The panels in Fig.~\ref{FIG9}(b) present the optical linewidth that arises from the electric and magnetic noise due to reverse leakage current [i.e., Eqs. (\ref{eq_lkg_17}) and (\ref{eq_lkg_24})] as a function of $x_{\text{def}}-D$ (separation between the spin center and the volume containing surface impurities){, as displayed in Fig.~\ref{FIG1}(a)}. We plot these variables for the same reverse bias voltages and initial diode parameters as in Fig.~\ref{FIG9}(a). In our simulations, the resulting linewidths are quite small {$(<10^{-12})$~MHz} for the case of magnetic Johnson noise, even for large reverse voltages such as $-1100$ V. This is a consequence of the small carrier density $(<10^8)$~cm$^{-3}$.
On the contrary, the charge noise due to the leakage current reaches kHz-MHz levels for spin centers near the surface, making it comparable to the linewidth resulting from majority carrier noise [i.e., Fig.~\ref{FIG3}(d)]. This linewidth decays rapidly with depth, as displayed in Fig.~\ref{FIG9}(b), mirroring the reduction in effective defect density $n_{\text{eff}}$ far from the interface shown in Fig.~\ref{FIG9}(a). Therefore, provided that the spin defects in 4H-SiC diodes are located at sufficiently large distances away from the surface (e.g., $x_{\text{def}}>100$ nm), the electromagnetic noise associated with reverse leakage current does not give rise to a linewidth comparable to that produced by majority carriers, even for large reverse bias voltages. As a result, we conclude that the leakage current is not a factor that constrains the optimization of the diode's parameters. Nevertheless, we stress that these results depend strongly on the density of surface trap states, which may vary from diode to diode.  

\section{Equations for $\delta \textbf{E}$}
\label{AppC1}

Here, we present the equations used to quantify the linewidth due to majority carriers in the non-depleted regions of our diode. We use the formalism developed in Ref. \cite{Candido1}, where the optical linewidth $\Gamma$ is modeled as the standard
deviation of the electric field produced by a fluctuating electric
dipole density. The electric field fluctuations are calculated via $\delta\mathbf{E}^{2}\equiv\int d^{3}r\,\rho\left(\mathbf{r}\right)\mathbf{E}_{d}^{2}\left(\mathbf{r}-\mathbf{r}_\text{def}\right)$, with $\mathbf{E}_{d}$ as the dipole electric field and $\mathbf{r}_\text{def}$ the spin center position and $\rho\left(\mathbf{r}\right)$ as the dipole density.

For a $pnn^+$ diode, the spin center is placed within the lightly-doped $n$-region.
As we increase the magnitude of the bias voltage $V$, the lightly-doped region progressively becomes more depleted. Additionally, the spin center experiences electric noise coming from both the bulk of the $p$ and $n^+$ regions. As a result,
for any spin center coordinate $z_{\text{def}}$ such that $0<z_{\text{def}}<d$, with $d$ being the length of the intrinsic region [Fig. \ref{FIG1}(a)], the defect will receive noise contributions coming from both the $p$ $(|\delta\mathbf{E}_{p}|)$ and $n^+$ regions $(|\delta\mathbf{E}_{n^{+}}|)$,
as well as from either the non-depleted $(|\delta\mathbf{E'}_{n}|)$ or the
depleted lightly-doped $n$-region $(|\delta\mathbf{E}_{n}|)$, depending on the defect's location $z_{\text{def}}$. Thus,
\begin{equation}
|\delta\mathbf{E}(z)|^2 = 
\begin{cases}
|\delta\mathbf{E}_{n^{+}}|^2 
+ |\delta\mathbf{E}_{p}|^2 
+ |\delta\mathbf{E}_{n}|^2 
& 
\begin{aligned}
& 0 < z < \tilde{d}_{n},
\end{aligned}
\\[4pt]
|\delta\mathbf{E}_{n^{+}}|^2 
+ |\delta\mathbf{E}_{p}|^2 
+ |\delta\mathbf{E'}_{n}|^2 
&
\begin{aligned}
& \tilde{d}_{n} < z_{\text{def}} < d,
\end{aligned}
\end{cases}
\label{EQC1_1}
\end{equation}
where $\tilde{d_{n}}$ is the depletion length of the diode's intrinsic $n$-region,
and
\begin{align}
|\delta\mathbf{E}_{n^{+}}| &= \frac{e}{4\pi\epsilon}d_{i,n^{+}}\sqrt{\frac{\pi}{3\Omega_{n_{+}}}} \notag \\
& \quad \times \sqrt{\left(\frac{1}{\left[d+d_{n}-z_{\text{def}}\right]^{3}} - \frac{1}{\left[d+d_{r}-z_{\text{def}}\right]^{3}}\right)} \label{EQC1_2}, \\
|\delta\mathbf{E}_{p}| &= \frac{e}{4\pi\epsilon}d_{i,p}\sqrt{\frac{\pi}{3\Omega_{p}}} \notag \\
& \quad \times \sqrt{\left(\frac{1}{\left[d_{p}+z_{\text{def}}\right]^{3}} - \frac{1}{\left[d_{l}+z_{\text{def}}\right]^{3}}\right)} \label{EQC1_3}, \\
|\delta\mathbf{E'}_{n}| &= \frac{e}{\sqrt{2}\pi\epsilon}n^{2/3}\left(z_{\text{def}}\right) \simeq \frac{e}{\sqrt{2}\pi\epsilon}\left(N_{N}\right)^{2/3} \label{EQC1_4}, \\
|\delta\mathbf{E}_{n}| &= \frac{e}{4\pi\epsilon}d_{i,n}\sqrt{\frac{\pi}{3\Omega_{n}}} \notag \\
& \quad \times \sqrt{\left(\frac{1}{\left[\tilde{d_{n}}-z_{\text{def}}\right]^{3}} - \frac{1}{\left[d-z_{\text{def}}\right]^{3}}\right)} \label{EQC1_5}.
\end{align}
with $\Omega_n=1/N_n$, $\Omega_p=1/N_a$ and $\Omega_{n+}=1/N_d$, and inter-dopant separation $d_i \equiv \Omega_i^{1/3}$ for each region. Here, $d_l$ and $d_r$ represent the physical lengths of the $p$ and $n^+$ layers, while $d_p$ and $d_n$ denote the corresponding depletion widths within those regions.

\section{Noise Optimization}
\label{AppD}

Here, we outline our algorithm used to find the sets of design parameters \( \mathbf{p} \) that minimize the linewidth \( \Gamma \) produced by majority carriers in our diode, which is the dominant contribution to decoherence of our spin centers. To achieve this, we use a scaled gradient descent method \cite{SGDO1,SGDO2} that accounts for relevant physical constraints such as dielectric breakdown voltage, upper/lower thresholds on doping densities and diode dimensions.

\subsection{Scaled Gradient Descent}

 
Let \( f(\mathbf{p}) = \alpha \hat{f}(D^{-1}\mathbf{p}) \) be a multi-variable function, where \( D \) is a diagonal matrix, \( D_{ii} > 0 \) the scale of each component of the vector of design parameters $\mathbf{p}$, and \( \alpha \) the scale for values of \( f \), with $\hat{f}$ being a well-scaled function. Our assumption
 is that the standard gradient descent algorithm performs well on $\hat{f}$, but that the computed function to be optimized is $f$, which we refer to as the target function. We update $\mathbf{p}$ in the negative gradient direction to decrease the numerical value of the target function as we progress through the iterations:
\begin{equation}
\hat{\mathbf{q}} \rightarrow \hat{\mathbf{p}} - s_k \nabla \hat{f}(\hat{\mathbf{p}}), 
\label{EQ10}
\end{equation}

\noindent where $\hat{\mathbf{p}}\equiv D^{-1}\mathbf{p}$ and $\hat{\mathbf{q}}\equiv D^{-1}\mathbf{q}$. Here, $s_k$ is the learning rate, whose role is to dictate how quickly the iterations move in the direction of the negative gradient. For any function $g\,(z)$, one can write \( g(z) = \hat{g}\,(Az) \), where $A$ is a diagonal matrix. Then, $\nabla g(z) = A^T \nabla \hat{g}(Az)$. Since $D$ is diagonal, $\nabla f\left(\mathbf{p}\right)=\alpha D^{-1} \nabla \hat{f}\left(D^{-1}\mathbf{p}\right)$, and so $\nabla \hat{f}\left(D^{-1}\mathbf{p}\right)=\alpha^{-1} D \nabla f\left(\mathbf{p}\right)$. The gradient step defined in Eq. (\ref{EQ10}) becomes

\begin{equation*}
\ D^{-1}\mathbf{q} \rightarrow D^{-1}\mathbf{p} - s_k \alpha^{-1} D^{T} \nabla f(\mathbf{p}), \
\end{equation*}
\begin{equation}
\mathbf{q} \rightarrow \mathbf{p}  - s_k \alpha^{-1} D^{2} \nabla f(\mathbf{q}).
\label{EQ11}
\end{equation}


\begin{figure*}[ht]
    \centering
    \includegraphics[width=\textwidth]{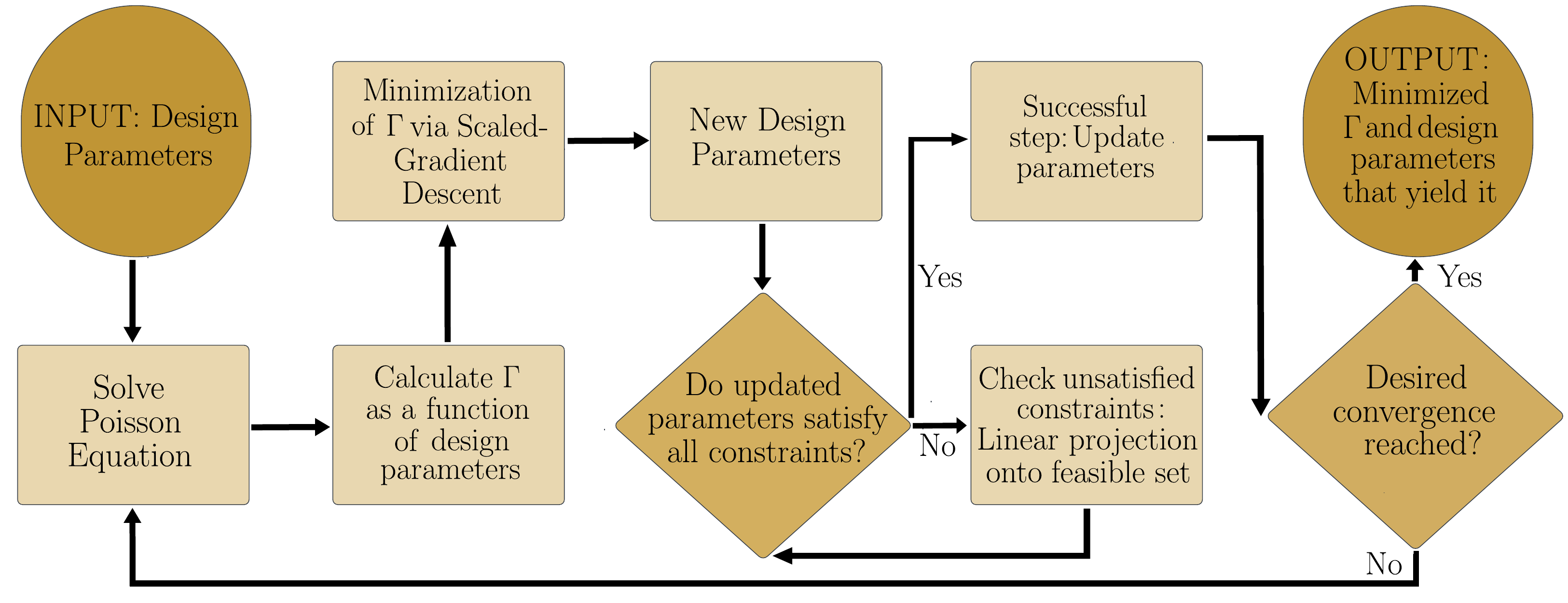}
    \caption{Flow chart of our optimization algorithm. For a given set of initial design parameters, the optical linewidth is calculated by means of solving the diode's Poisson equation and determining its charge carrier concentrations at thermal equilibrium. Then, the scaled-gradient descent method is applied to minimize the linewidth target function. At each iteration, it is necessary to satisfy the diode's physical constraints. Once convergence is achieved, the algorithm yields the final diode design parameters and the minimized optical linewidth function.}
    \label{FIG10}
\end{figure*}

\subsection{Constraint Optimization}

Eq. (\ref{EQ11}) needs to be modified to account for the physical constraints inherent to our optimization problem. For example, we want to avoid dielectric breakdown of the diode as well as other constraints on the design parameters $\mathbf{p}$. The physical constraints are functions of $\mathbf{p}$, and are represented by the inequalities $h_j(\mathbf{p}) \geq 0$, with $j = 1,2,...,m$,  with $m$ as the number of physical constraints of our system.  Hence, the problem becomes to solve
\begin{equation}
    \min_{\mathbf{p}} \; f(\mathbf{p}) \;
\text{subject to} \quad h_j(\mathbf{p}) \ge 0, \; j=1,\dots,m.
\end{equation}
The constraint functions also depend on the design parameters, it is crucial to also scale the constraint functions, so we write
\begin{equation*}
h_j(\mathbf{p}) = \beta_j\hat{h_j}(\beta^{-1}\mathbf{p}), 
\end{equation*}

\noindent where $\beta_{j} > 0 $ is the scale of the constraint function $h_j$. Here, $h_j(\mathbf{p})$ is assumed to be a well-scaled function of $\hat{\mathbf{p}}=D^{-1}\mathbf{p}$. Since reducing $f(\mathbf{p})$ may lead to infeasible designs (i.e., $h_j(\mathbf{p}) < 0$ for some $j$), it is important to correct $\mathbf{p}$ so that the constraints $h_j(\mathbf{p}) \geq 0$ are satisfied. A first-order linear projection of $\mathbf{q}$ onto the feasible set is achieved by enforcing $h_j(\mathbf{q}+\mathbf{\delta q})\approx h_j(\mathbf{q})+\nabla h_j(\mathbf{q}) \mathbf{\delta q} = 0$, $\forall\,\delta q_i\ll\beta_{ii}$. Thus, we seek $\mathbf{q}$ that minimizes $|D ^{-1} \mathbf{q}|$, where $h_j(\mathbf{q})+\nabla h_j(\mathbf{q}) \mathbf{\delta q} = 0$. Solving this constrained optimization problem gives
 
\begin{equation}
\delta \mathbf{q} = -\frac{h_j(\mathbf{q})}{\lvert D \nabla h_j(\mathbf{q}) \rvert^2} D^2 \nabla h_j(\mathbf{q}).
\label{EQ12}
\end{equation}

\noindent The update for $\mathbf{q}$ to correct for a violation of the constraints $h_j(\mathbf{p}) \geq 0$ would then be

\begin{equation}
\mathbf{q} \rightarrow \mathbf{p}  - \gamma\frac{h_j(\mathbf{q})}{|D\nabla h_j(\mathbf{q})|^2}  D^{2} \nabla h_j(\mathbf{q}),
\label{EQ13}
\end{equation}

\noindent where $0<\gamma<1$ is a constant that helps stabilize the projection step.

\subsection{Optimization Algorithm}

 Here, we present the formalism on which each step of our algorithm is based, which is illustrated by Fig. \ref{FIG10}.  The first step is to solve the diode's Poisson equation [Eq. (\ref{EQ8_modified})] to determine the free carrier density profile, and then use Eq. (\ref{lnwdth14}) together with Eqs. (\ref{EQC1_2})-(\ref{EQC1_5}) to calculate the corresponding optical linewidth $\Gamma(\mathbf{p},z_{\text{def}})$. Then, we apply the scaled gradient descent procedure to find local minima of $\Gamma$ within the parameter space defined by $\textbf{p}$. Finally, we determine the minimum of $\Gamma$ with respect to $z_{\text{def}}$ corresponding to the design parameters updated via gradient descent.

The gradient descent scheme needs to follow a structure such that it performs the optimization while 1) keeping the value of the target function within the feasible set defined by our constraint functions $h_j$ {(i.e. $h_j(\mathbf{p})\geq0$),} and 2) making sure that the value of the target function is decreasing overall. The first condition can be expressed as follows. Let the $L$-Infinity norm of a vector $\mathbf{x}$ be defined as $|x|_\infty = \max_{i=1,2,\ldots,n} |x_i|$. Then, satisfying the constraints of the optimization problem requires that, for any gradient descent update,
\begin{equation}
|\beta^{-1} \mathbf{h(q)}_-|_{\infty}=0,
\label{EQ14}
\end{equation}
where $\mathbf{h(q)}_- = [\min(0, h_j(\mathbf{q})), \; j = 1,2,\ldots,m]$. Thus, $|\beta^{-1} \mathbf{h(q)}_-|_{\infty}$ gives the worst constraint violation after scaling each constraint function. In a constrained optimization problem, it is necessary to balance both reducing the objective function and satisfying the constraints on the design parameters. To accomplish this, we introduce a merit function  that explicitly shows the balance between these two aims:
\begin{equation}
m(\mathbf{p}) = f(\mathbf{p}) + \alpha M \lvert \beta^{-1} h(\mathbf{p})_- \rvert_\infty,
\label{EQ15}
\end{equation}
with $M>0$ and $\alpha>0$ as proportionality constants. To ensure convergence across varying landscapes, we employ a non-monotonic descent criterion \cite{non_monotone}, where a proposed step $\mathbf{q}$ is accepted if its merit function value satisfies 
\begin{equation}
    m(\mathbf{q}) < \langle m(\mathbf{p})\rangle_k ,
    \label{eq:non_monotone}
\end{equation}
with $\langle m(\mathbf{p})\rangle_k$ representing the moving average of the merit function over the previous $k$ iterations. As a result, every time that Eq. (\ref{EQ14}) is not satisfied, i.e., 
\begin{equation}
|\beta^{-1} \mathbf{h(q)}_-|_{\infty}>0, 
\label{EQ16}
\end{equation}
we recalculate the updated design parameters using Eq. (\ref{EQ13}) for all constraints $h_j$ that are not obeyed, and repeat this up to a maximum number of iterations $n_\text{max}$, or until Eq. (\ref{EQ14}) is obeyed. In our work, we use $n_\text{max}=100$. Then, we test if there is a reduction in the merit function \textcolor{blue}{given by} Eq. (\ref{eq:non_monotone}). If Eq. (\ref{eq:non_monotone}) is satisfied, $\mathbf{q}$ becomes our updated set of design parameters and we proceed to the next iteration without modifying the learning rate $s_k$. Otherwise, $s_k$ is decreased so that the upcoming iterations produce parameters $\mathbf{q}$ that reduce the merit function. In our algorithm, the learning rate self-adjusts so that it becomes larger when the merit function decreases, and goes down when the merit function increases. i.e., we start out with $s_k=s_{k,\text{min}}$ and allow the learning rate to increase only up to $s_k=s_{k,\text{max}}$. This procedure is related to the use of adaptive step sizes \cite{adaptiveSIZE1,adaptiveSIZE2} to help the evolution in parameter space be smooth. Provided that $M$ is sufficiently large, the minimization of the merit function leads to solving the constrained optimization problem. If the objective function $f(\mathbf{p})$ keeps decreasing while the constraint violations $|\beta^{-1} \mathbf{h(p)}_-|_{\infty}$ continue to increase, $M$ should be increased, for example, by doubling its value. In our simulations, we sue $s_{k,\text{min}}=5\times10^{-5}$, $s_{k,\text{max}}=150$, and $M=\alpha=2$.

In our algorithm, we calculate partial derivatives via finite differences, which can introduce significant errors, primarily due to the truncation of higher-order Taylor series terms\cite{trunc_error}. This often leads to stagnation and oscillations in the optimization trajectory near local minima \cite{trunc_error}. To mitigate this, we employ a search method, consisting of averaging the partial derivatives of the last $l$ iterations, so that we obtain a non-negligible variation in the target function when standard finite differences fail to capture physical trends near local minima. The aforementioned strategy is related to pattern search \cite{search_opt,search_opt2} methods, which are standard techniques in optimization protocols. Finally, we define practical convergence as occurring when the fractional change in the target function falls below a threshold $\chi$ for $n$ consecutive iterations: $|f\,(\mathbf{q}) - f\,(\mathbf{p})|/f\,(\mathbf{p}) < \chi$. Additionally, if the algorithm detects oscillations about the smallest achieved value with an amplitude less than $\chi$ over $n$ iterations, convergence is also declared. For our simulations, we choose $n=30$ and $\chi=10^{-5}$.


\nocite{*}


\bibliography{apssamp.bib}

\end{document}